\documentclass[%
 reprint,
 amsmath,amssymb,
 aps,
pra,
]{revtex4-2}

\usepackage{subfigure}
\subfiguretopcaptrue
\usepackage{graphicx}
\usepackage[dvipsnames]{xcolor}
\usepackage{dcolumn}
\usepackage{bm}
\usepackage{amsmath,amsfonts,amssymb}
\DeclareMathOperator{\sech}{sech}

\begin{document}

\title{Efficient exciton generation in a semiconductor quantum dot-metal nanoparticle composite structure using conventional chirped pulses}

\author{Dionisis Stefanatos}
\email{dionisis@post.harvard.edu}

\author{Athanasios Smponias}

\author{Ioannis Thanopulos}

\author{Emmanuel Paspalakis}

\affiliation{Materials Science Department, School of Natural Sciences, University of Patras, Patras 26504, Greece}

\date{\today}

\begin{abstract}
We consider a nanostructure consisting of a semiconductor quantum dot coupled to a metal nanoparticle, and show with numerical simulations that the exciton state of the quantum dot can be robustly generated from the ground state even for small interparticle distances, using conventional chirped pulses with Gaussian and hyperbolic secant envelopes. The asymmetry observed in the final exciton population with respect to the chirp sign of the applied pulses is explained using the nonlinear density matrix equations describing the system, and is attributed to the real part of the parameter emerging from the interaction between excitons in the quantum dot and plasmons in the metal nanoparticle. The simplicity of the conventional chirped pulses, which can also be easily implemented in the laboratory, make the proposed robust quantum control scheme potentially useful
for the implementation of ultrafast nanoswitches and quantum information processing tasks with semiconductor quantum dots.
\end{abstract}

\maketitle
\bibliographystyle{apsrev}

\section{Introduction}
An intense field of research is devoted to studies regarding the optical properties of complex systems composed of plasmonic nanostructures coupled to quantum entities like molecules or semiconductor quantum dots (SQDs) \cite{Review}. When the quantum part of these composite nanosystems is coherently controlled, they behave as active nanophotonic structures and are expected to have important applications in many fields, including nanotechnology and modern quantum technologies. For example, it has been found that a composite structure which consists of a semiconductor quantum dot (SQD) and a metal nanoparticle (MNP), is more efficient than a quantum dot alone for optical phenomena like the creation of single photons on demand \cite{single1,single2} and polarization-entangled photons \cite{entangled}. The coupled SQD-MNP nanostructure serves also as the basic system for the plasmonic nanolaser (spaser) \cite{spaser1,spaser2}. In order to exploit the advantages offered by the coupled SQD-MNP system for these important quantum technology applications, a crucial problem is the efficient controlled population transfer from the ground to the exciton state of the quantum dot, in the presence of the nanoparticle.
This important problem has been explored in a series of studies
\cite{Cheng07a,Sadeghi09a,Sadeghi09b,Malyshev11a,Malyshev13a,Carreno18a,Paspalakis13a,Sadeghi10a,Lee15a,McMillan16a,Anton12a,Qi19a}, with emphasis put to the effect of the interparticle distance.

More precisely, by studying a nanostructure containing a CdTe SQD and a rodlike Au MNP, it was discovered that the period of Rabi oscillations exhibited by the exciton population is modified with the interparticle distance \cite{Cheng07a}, an effect that was later associated with the development of plasmonic metaresonances \cite{Sadeghi09a,Sadeghi09b}. Moreover, this last phenomenon was linked to optical bistability that may take place in a SQD-MNP nanosystem \cite{Malyshev11a,Malyshev13a,Carreno18a,Paspalakis14a}. It was also shown that properly tailored pulses with hyperbolic secant envelope can achieve high fidelity exciton state preparation in a SQD coupled to a spherical MNP \cite{Paspalakis13a}. The application of ultrashort pulse trains or amplitude modulated laser pulses to the same system, results in distance-dependent modulation of the exciton population in the SQD, a phenomenon which can be exploited for the implementation of efficient nanoswitches \cite{Sadeghi10a,Lee15a}. In another study considering a three-level V-type SQD, the MNP was exploited in order to obtain selective population transfer to one of the exciton states by applying resonant fields \cite{Anton12a}. In a related work, optimal control was used for the effective transfer of population between the two lower states of a $\Lambda$-type SQD placed close to a spherical MNP \cite{Qi19a}.

In the majority of the previously discussed works, resonant methods have been employed for the preparation of the SQD exciton state in the presence of MNP. The main advantage of these methods is the fast and with high fidelity population transfer to the exciton state, something which occurs only when using finely tuned pulse amplitudes and widths. The efficiency of resonant methods is rather sensitive to changes in the characteristics of the applied fields. A way to overcome these drawbacks and obtain robust population inversion in a two-level system is to use adiabatic methods \cite{Vitanov01a,Goswami03a} or the closely related shortcuts to adiabaticity \cite{Shortcutreview}, where the latter are essentially accelerated versions of the former, while both are implemented using chirped pulses. In our recent work we used the shortcut method of transitionless quantum driving and showed that efficient preparation of the exciton state in  SQD-MNP system can be accomplished \cite{Sbonias21}.

In the current article we use conventional chirped pulses with Gaussian or hyperbolic secant envelopes and linear or hyperbolic tangent chirp, respectively, and demonstrate that they also can robustly prepare the exciton state in a SQD coupled to a MNP. The reason for considering such pulses is their simplicity and easiness to implement in the laboratory, compared to the more sophisticated chirp and envelopes needed by the shortcut pulses. Note that these type of pulses have been used for the efficient preparation of the exciton \cite{Yu11a,Simon11,Luker12,Wei14a,Mathew14a,Kaldewey17b,Mukherjee20,Ramachandran21} and biexciton \cite{Hui08a,Axt13a,Amand13a,Kaldewey17a} states in a SQD in the absence of a MNP. In the present study involving the coupled SQD-MNP system and for the employed chirped pulses we observe an asymmetry in the final exciton population, which depends on the sign of the chirp parameter, positive or negative. In the nonlinear density matrix equations describing the coupled SQD-MNP system \cite{Cheng07a,Sadeghi09a,Sadeghi09b,Malyshev11a,Malyshev13a,Carreno18a,Paspalakis13a,Sadeghi10a,Lee15a,McMillan16a,Wang06a,Yan08a,Hatef12a,Li12a,Kosionis12a,Kosionis13a,Sadeghi17a,Miri18a,Kosionis18a,Hapuarachchi18a,Kosionis19a}, we explicitly identify the symmetry breaking term as the real part of the nonlinearity self-interaction parameter, where the latter is due to the dipolar exciton-plasmon interaction \cite{Wang06a,Yan08a}. Using the familiar two-level system terminology, this term corresponds to an effective ``longitudinal" field which breaks the $z$-symmetry in the Bloch sphere, affecting differently the ``longitudinal" field associated with positive and negative chirp parameter.

The present work has the following structure. In section \ref{sec:system} we provide the nonlinear density matrix equations which describe the interaction of the SQD-MNP system with the applied electromagnetic field. In section \ref{sec:pulses} we describe the applied chirped pulses and also discuss the symmetry breaking in the equations for opposite chirp parameters. In section \ref{sec:results} we provide numerically obtained fidelity diagrams of the exciton state final population, for various pulse durations and SQD-MNP distances. Section \ref{sec:con} summarizes the results of the current research.

\section{Coupled semiconductor quantum dot-Metal nanoparticle model}

\label{sec:system}

\begin{figure}[t]
\centering
\includegraphics[width=0.85\linewidth]{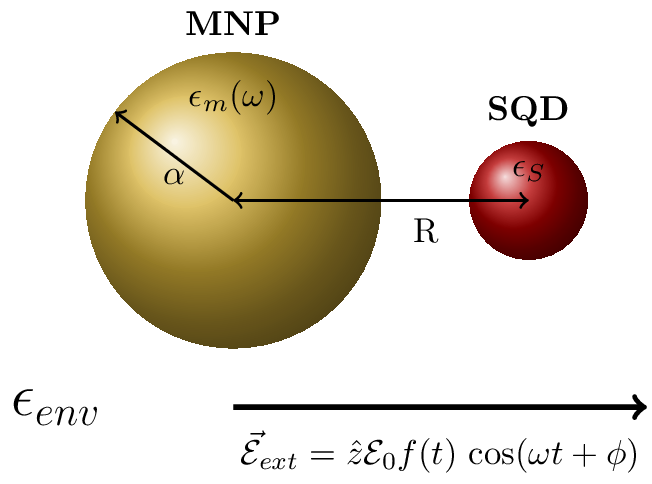} \label{fig1}
\caption{Spherical metal nanoparticle with radius $\alpha$ and dielectric constant $\varepsilon_m(\omega)$ coupled to a semiconductor quantum dot with radius $b\ll \alpha$ and dielectric constant $\varepsilon_s$. $R$ denotes the interparticle distance, $\varepsilon_{env}$ the dielectric constant of the surrounding, and $\mathcal{\vec{E}}(t)$ the externally applied field.}
\label{fig:nanostructure}
\end{figure}

The system under consideration is displayed in Fig. \ref{fig:nanostructure} and consists of a classical spherical MNP with radius $\alpha$ and dielectric function $\varepsilon_{m}(\omega)$, and a SQD with dielectric constant $\varepsilon_{s}$ which is modeled as a two-level system, with states $|0\rangle$ and $|1\rangle$ corresponding to the ground and single exciton states, respectively, an approximation used in several previous works \cite{Cheng07a,Sadeghi09a,Sadeghi09b,Malyshev11a,Malyshev13a,Carreno18a,Paspalakis13a,Sadeghi10a,Lee15a,McMillan16a,Wang06a,Yan08a,Hatef12a,Li12a,Kosionis12a,Kosionis13a,Sadeghi17a,Miri18a,Kosionis18a,Hapuarachchi18a,Kosionis19a}. The two components of the nanosystem are embedded in environment with dielectric constant $\varepsilon_{env}$, with their centers separated by a distance $R$.

A linearly polarized external electric field $\mathcal{\vec{E}}(t) = \hat{z}\mathcal{E}_{0} f(t)\cos[\omega t + \phi(t)]$ is applied to the nanostructure, where $\mathcal{E}_{0}$ is the electric field amplitude, $f(t)$ is the dimensionless pulse envelope, $\omega$ is the angular frequency and $\phi(t)$ is the time-dependent phase. The Hamiltonian describing the interaction is
\begin{equation}
\mathcal{H} = \hbar\omega_{0}|1\rangle \langle 1| -\mu \mathcal{E}_{SQD}(t)\left(|0\rangle \langle 1|+|1\rangle \langle 0|\right) \, ,\label{hamilt}
\end{equation}
where $\hbar\omega_0$ is the exciton state energy, $\mu$ is the dipole moment for the ground to exciton transition in the SQD, which without loss of generality is assumed to be real, while $\mathcal{E}_{SQD}$ expresses the total electric field inside the SQD. This latter quantity is composed of two terms in the dipole approximation, one corresponding to the applied external field and another to the field induced in the SQD by the polarization of the MNP. If we split the positive and negative frequency contributions, since they present distinct time responses, we get for $\mathcal{E}_{SQD}$ the expression \cite{Wang06a,Yan08a,Artuso10a}
\begin{eqnarray}
\mathcal{E}_{SQD}(t) &=& \frac{\hbar}{\mu}\bigg[\left(\frac{\Omega(t)}{2} + G \sigma(t)\right)e^{-i[\omega t + \phi(t)]} \nonumber \\
&+& \left(\frac{\Omega^{*}(t)}{2} + G^{*} \sigma^{*}(t)\right)e^{i[\omega t + \phi(t)]} \bigg] \, ,
\end{eqnarray}
where $\displaystyle{\sigma(t) = \rho_{10}(t)e^{i[\omega t + \phi(t)]}}$ is the slowly varying off-diagonal density matrix element of the SQD. Additionally, $\Omega(t)$ denotes the time-dependent complex Rabi frequency given by \cite{Wang06a,Yan08a,Artuso10a}
\begin{equation}
\Omega(t) = \Omega_{0}f(t), \quad \Omega_{0} = \frac{\mu \mathcal{E}_{0}}{\hbar \varepsilon_{effS}}\left(1 + \frac{s_{a}\gamma_{1}\alpha^{3}}{R^{3}}\right) \, , \label{rabif}
\end{equation}
while $G$ is the self-interaction parameter defined as \cite{Yan08a}
\begin{equation}
G =  \sum^{N}_{n=1}\frac{1}{4\pi\varepsilon_{env}}\frac{(n+1)^2\gamma_{n}\alpha^{2n+1}\mu^{2}}{\hbar \varepsilon^{2}_{effS} R^{2n+4}} \, . \label{gpar}
\end{equation}
In the above formulas, the quantity $\displaystyle{\varepsilon_{effS} = \frac{2\varepsilon_{env}+\varepsilon_{S}}{3\varepsilon_{env}}}$ represents the SQD effective dielectric constant, $\displaystyle{\gamma_{n} = \frac{\varepsilon_{m}(\omega) - \varepsilon_{env}}{\varepsilon_{m}(\omega)+(n+1)\varepsilon_{env}/n}}$ with $n = 1, 2, 3, \dots$, while $s_{a} = 2$ since the external field is parallel to the centre line of the system ($z$-axis).

\begin{figure}[t]
 \centering
		\begin{tabular}{c}
        \subfigure[$\ $]{
	            \label{fig:Re_G}
	            \includegraphics[width=0.85\linewidth]{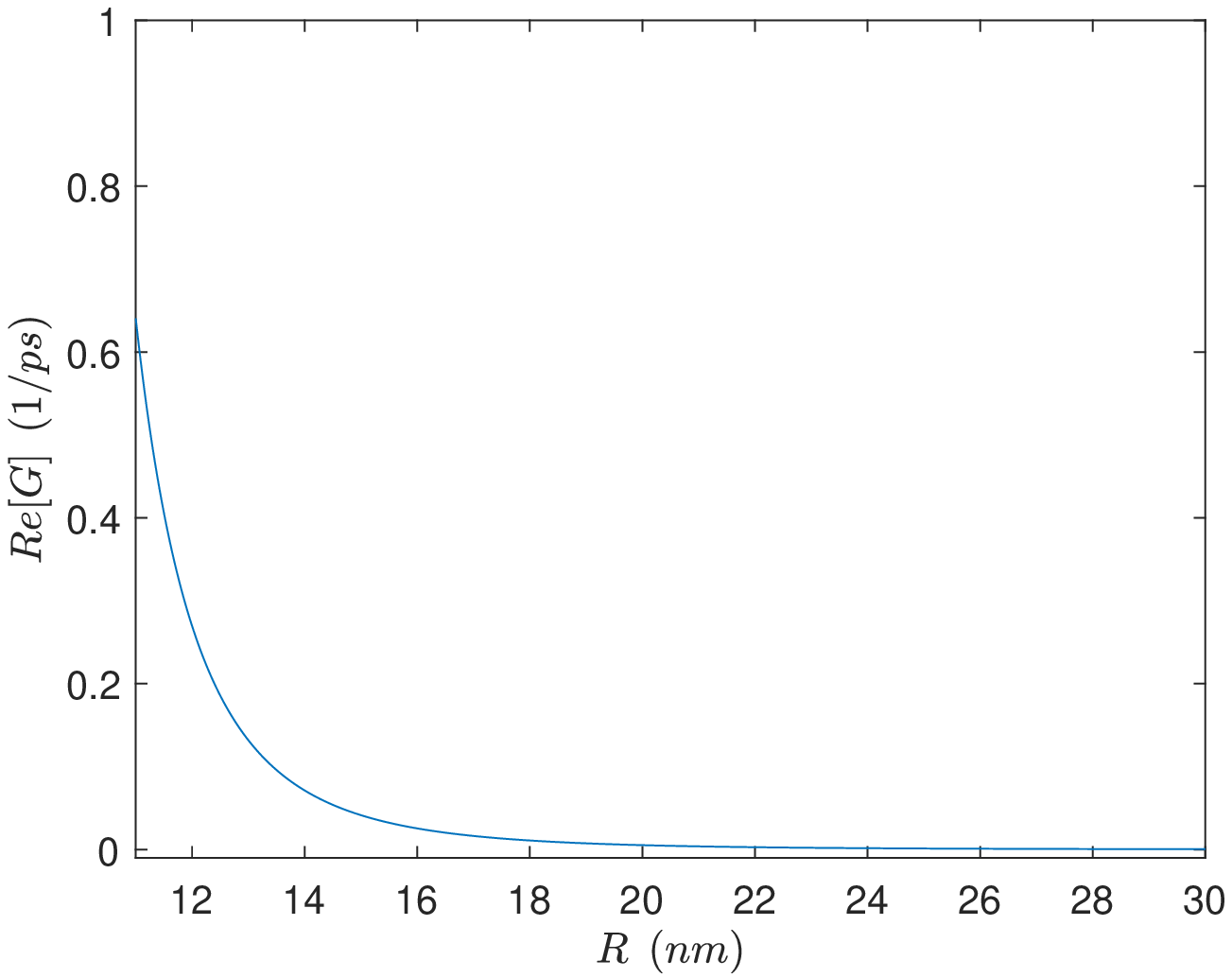}} \\
        \subfigure[$\ $]{
	            \label{fig:Im_G}
	            \includegraphics[width=0.85\linewidth]{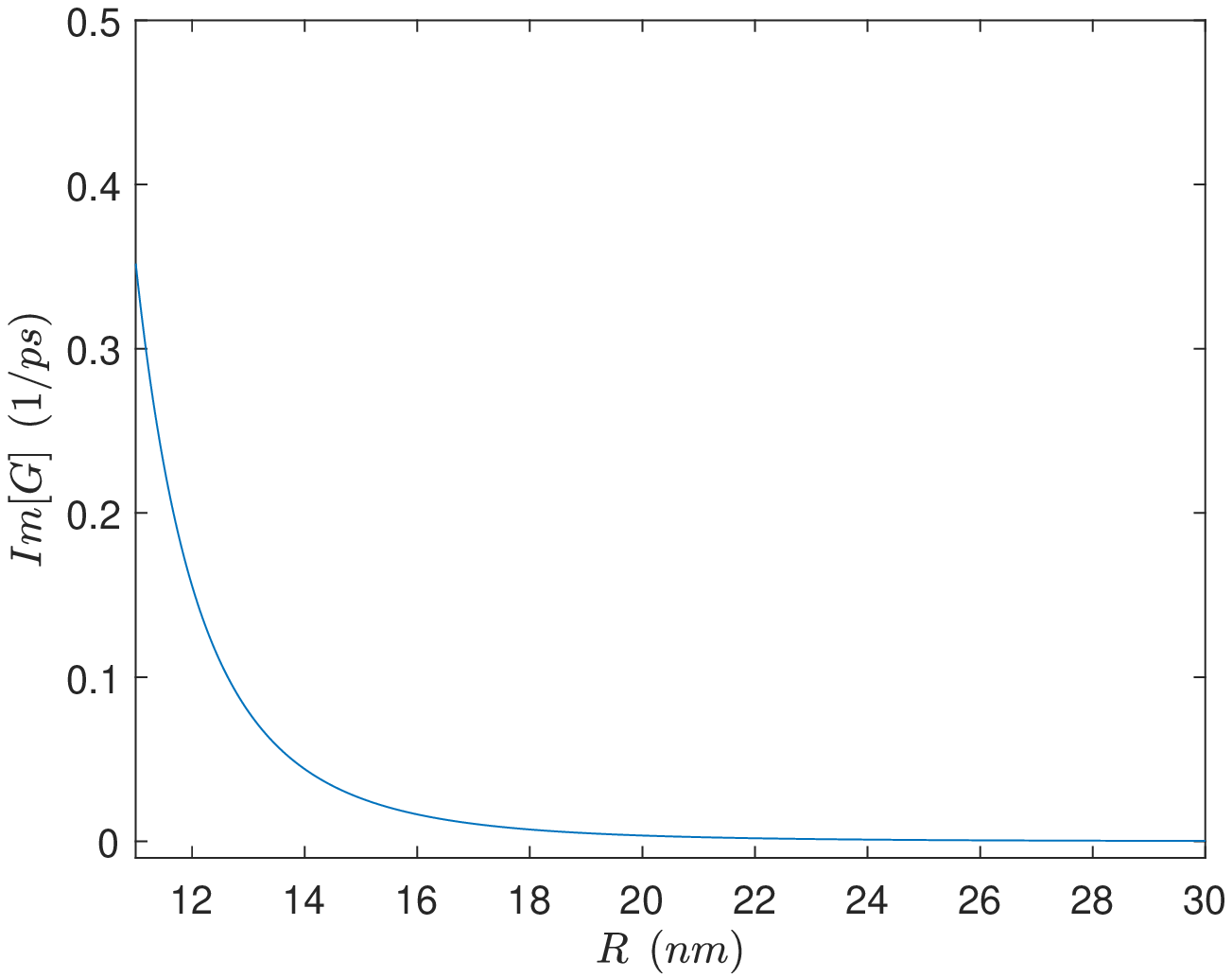}}
		\end{tabular}
\caption{(Color online) Real (a) and imaginary (b) parts of the nonlinear self-interaction parameter $G$, as a function of the interparticle distance.}
\label{fig:G}
\end{figure}

The two parts of the complex Rabi amplitude correspond to the applied field and the field induced by the polarization of the MNP (caused also by the external field). If we write $\Omega_0=|\Omega_0|e^{i\beta}$, where $\beta$ is the corresponding phase, then the complex Rabi frequency can be expressed as $\Omega(t)=|\Omega(t)|e^{i\beta}$, where $|\Omega(t)|=|\Omega_0|f(t)$, a relation which we shall use in the following section. The self-interaction parameter $G$ emerges from the interaction between SQD excitons and MNP plasmons \cite{Wang06a,Cheng07a,Yan08a}. Particularly, the external field induces a dipole on the SQD, proportional to the off-diagonal elements of the density matrix, which also induces a dipole on the MNP, which subsequently interacts with the SQD dipole \cite{Wang06a,Artuso10a}, with this last interaction being included in Hamiltonian (\ref{hamilt}). In Eq. (\ref{gpar}) we actually consider that the SQD dipole induces multipole moments on the MNP \cite{Yan08a,Artuso11a}, and use the value $N=20$ in the subsequent calculations in order to get converging results. In Fig. \ref{fig:G} we display the real and imaginary parts of $G$ as a function of the interparticle distance, for the parameter values later used in section \ref{sec:results}.

Starting from Hamiltonian (\ref{hamilt}) we find a set of equations for the population difference between the ground and the single exciton states, $\Delta(t) = \rho_{00}(t) - \rho_{11}(t)$, and the slowly varying off-diagonal matrix element $\displaystyle{\sigma(t) = \rho_{10}(t)e^{i[\omega t + \phi(t)]}}$,
\begin{subequations}
\label{system}
\begin{eqnarray}
\dot{\Delta}(t)&=&i\Omega^*(t)\sigma(t)-i\Omega(t)\sigma^*(t)+4G_I\sigma(t)\sigma^*(t)\nonumber\\
&&-\frac{\Delta(t)-1}{T_1}, \label{delta}\\
\dot{\sigma}(t)&=&i(\delta+\dot{\phi}(t))\sigma(t)+i\frac{\Omega(t)}{2}\Delta(t)+iG\Delta(t)\sigma(t)\nonumber\\
&&-\frac{\sigma(t)}{T_2}, \label{twol_sigma}
\end{eqnarray}
\end{subequations}
where $\delta = \omega - \omega_{0}$ is the detuning of the applied field and $G_{I}$ is the imaginary part of $G=G_{R}+iG_{I}$, while $T_{1}, T_{2}$ denote the relaxation times corresponding to spontaneous emission and dephasing in the SQD, respectively. Observe that nonlinear terms arise in the above equations due to parameter $G$. At $t=0$ the initial conditions starting from the ground state are $\Delta(0)=1, \sigma(0)=0$, while complete exciton preparation at the final time $t=t_f$ corresponds to the target value $\Delta(t_f)=-1$. The quantity that we are interested in is the final exciton population $\rho_{11}(t_f)=[1-\Delta(t_f)]/2$.

\section{Conventional chirped pulses and symmetry breaking}

\label{sec:pulses}

In this section we explain how chirped pulses can be used for the efficient preparation of the exciton state, as well as why opposite sign chirp parameters lead to different final fidelities, due to the symmetry breaking of the system caused by the presence of the MNP. Observe that for $G=0$ and $T_1, T_2\rightarrow\infty$, Eqs. (\ref{system}) reduce to the Bloch equations for the two-level system
\begin{equation}
\label{two_level}
i
\left(
\begin{array}{c}
\dot{a_1}(t)\\
\dot{a_2}(t)
\end{array}
\right)
=
\frac{1}{2}
\left(
\begin{array}{cc}
-\dot{\phi}(t) & \Omega(t)\\
\Omega^*(t) & \dot{\phi}(t)
\end{array}
\right)
\left(
\begin{array}{c}
a_1(t)\\
a_2(t)
\end{array}
\right),
\end{equation}
with the mapping $\Delta(t)=|a_1(t)|^2-|a_2(t)|^2$ and $\sigma(t)=a_1(t)a_2^*(t)$. The exciton state preparation from $\Delta(0)=1$ to $\Delta(t_f)=-1$ corresponds to inverting the population in this two-level system. The instantaneous eigenstates of the two-level system and the corresponding eigenvalues are
\begin{subequations}
\label{eigenvectors}
\begin{eqnarray}
|\psi_{+}(t)\rangle&=&
\left(\begin{array}{c}
    \cos{\frac{\theta(t)}{2}}\\
    \sin{\frac{\theta(t)}{2}}e^{-i\beta}
\end{array}\right),\label{plus}\\
|\psi_{-}(t)\rangle&=&
\left(\begin{array}{c}
    \sin\frac{\theta(t)}{2}\\
    -\cos{\frac{\theta(t)}{2}}e^{-i\beta}
\end{array}\right) \label{minus}
\end{eqnarray}
\end{subequations}
and
\begin{equation}\label{eigenvalues}
A_{\pm}(t)=\pm\frac{1}{2}\sqrt{\dot{\phi}^2(t)+|\Omega(t)|^2} \, ,
\end{equation}
where
\begin{equation}
\tan{\theta(t)}=\frac{|\Omega(t)|}{-\dot{\phi}(t)}
\end{equation}
while recall that $\Omega(t)=|\Omega(t)|e^{i\beta}$. If the applied electric field is selected such that the mixing angle is slowly varied from $\theta(0)=0$ to $\theta(t_f)=\pi$, then the population inversion takes place adiabatically along the eigenstate $|\psi_{+}(t)\rangle$. If $\theta(0)=\pi$ is slowly modified to $\theta(t_f)=0$, then the inversion occurs along the eigenstate $|\psi_{-}(t)\rangle$.

\begin{figure}[t]
\centering
\includegraphics[width=0.85\linewidth]{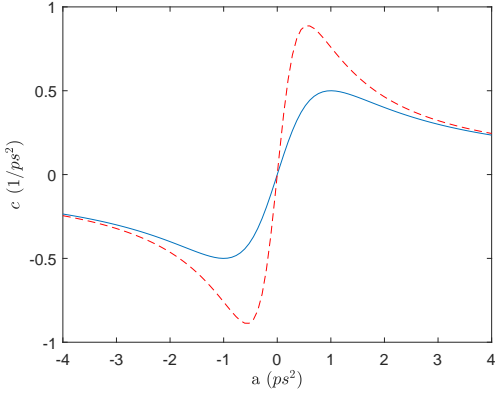}
\caption{(Color online) Chirp rate $c$ versus the chirp constant $a$ for two durations of the initial Gaussian pulse, $\tau_0=1$ ps (blue solid line) and $\tau_0=0.75$ ps (red dashed line).}
\label{fig:c}
\end{figure}

In order to achieve the desired population inversion, and thus the exciton state preparation, we will use initially linearly chirped Gaussian pulses. We explain briefly how such a pulse can be obtained when starting from a pulse with constant frequency and Gaussian profile
\begin{equation}\label{eq:Gauss}
f(t) = \exp \left[ - \frac{(t-t_0)^2}{2 \tau_0^2}\right],
\end{equation}
i.e.
\begin{equation}
\label{initial_pulse}
\mathcal{E} = \mathcal{E}_0 \exp \left[  - \frac{(t-t_0)^2}{2 \tau_0^2} \right] \cos \omega t,
\end{equation}
where
\begin{equation}
\mathcal{E}_0 = \frac{\hbar \epsilon_{effs}}{\mu}\frac{\Theta}{\sqrt{2\pi}\tau_0}
\end{equation}
is the amplitude and
\begin{equation}\label{eq:eurospalmou}
\Theta = \int_{-\infty}^{\infty} f(t) \mathop{dt},
\end{equation}
is the pulse area, which equals $\sqrt{2\pi}\tau_0$ for the profile \eqref{eq:Gauss}. If the pulse (\ref{initial_pulse}) passes through a chirp filter with chirp constant $a$, it is transformed to the pulse \cite{Saleh07,Luker12}
\begin{equation}
\label{G_chirped}
\mathcal{E}(t) = \frac{\hbar \epsilon_{effs}}{\mu}\frac{\Theta}{\sqrt{2\pi\tau_0 t_p}} \exp \left[  - \frac{(t-t_0)^2}{2 t_p^2} \right] \cos \left[ \omega t + \phi(t) \right],
\end{equation}
where its duration is modified from $\tau_0$ to \cite{Luker12,Malinovsky01}
\begin{equation}\label{eq:tpa}
t_p = \sqrt{\tau_0^2 + \frac{a^2}{\tau_0^2}}\,,
\end{equation}
while its frequency acquires a linear chirp
\begin{equation}
\label{chirp}
\dot{\phi}(t)=c(t-t_0),
\end{equation}
with rate \cite{Luker12,Malinovsky01}
\begin{equation}\label{eq:ca}
c = \frac{a}{a^2+\tau_0^4}.
\end{equation}
The chirp rate $c$ as a function of the chirp constant $a$ is displayed in Fig. \ref{fig:c}, for $\tau_0=1$ ps (blue solid line) and $\tau_0=0.75$ ps (red dashed line).
We will also use pulses with hyperbolic secant envelope
\begin{equation}
\label{sech}
\mathcal{E}(t) = \frac{\hbar \epsilon_{effs}}{\mu}\frac{\Theta}{\pi t_p} \sech\left(\frac{t-t_0}{t_p}\right) \cos \left[ \omega t + \phi(t) \right],
\end{equation}
and chirp
\begin{equation}
\label{tanh_chirp}
\dot{\phi}(t)=\frac{c}{t_p}\tanh{\left(\frac{t-t_0}{t_p}\right)}.
\end{equation}
Note that in this case the chirp parameter $c$ is dimensionless. For both types of pulses the final time is taken $t=2t_0$, where $t_0$ is large enough and defines the center of the pulse.

Observe that for $c>0$ ($a>0$ for Gaussian) the mixing angle changes from $0$ to $\pi$ for both types of pulses, thus the system evolves along $|\psi_{+}(t)\rangle$, while for $c<0$ ($a<0$ for Gaussian) it changes from $\pi$ to $0$ and the system evolves along $|\psi_{-}(t)\rangle$. For $G=0$ the two paths are equivalent but the presence of a $G=G_R+iG_I$ with $G_R\neq 0$ breaks this symmetry, as we immediately show. Let $\Delta(t), \sigma(t)$ be the solution of system (\ref{system}), with $\delta=0$ and ignoring relaxation
\begin{subequations}
\label{system1}
\begin{eqnarray}
\dot{\Delta}(t)&=&i\Omega^*(t)\sigma(t)-i\Omega(t)\sigma^*(t)+4G_I\sigma(t)\sigma^*(t), \label{delta1}\\
\dot{\sigma}(t)&=&i\dot{\phi}(t)\sigma(t)+i\frac{\Omega(t)}{2}\Delta(t)+iG\Delta(t)\sigma(t), \label{sigma1}
\end{eqnarray}
\end{subequations}
when starting from $\Delta(0)=1, \sigma(0)=0$ and with chirp $\dot{\phi}$ given by Eq. (\ref{chirp}).
Let also $\Delta'(t), \sigma'(t)$ be the solution when starting from the same initial conditions $\Delta'(0)=1, \sigma'(0)=0$ but with the opposite chirp $\dot{\phi}'=-\dot{\phi}$, i.e. when using $-c$ \,(or $-a$) in Eq. (\ref{chirp}). The primed variables satisfy the equations
\begin{subequations}
\label{system_prime}
\begin{eqnarray}
\dot{\Delta}'(t)&=&i\Omega^*(t)\sigma'(t)-i\Omega(t)\sigma'^*(t)+4G_I\sigma'(t)\sigma'^*(t), \label{delta_prime}\quad\quad\\
\dot{\sigma}'(t)&=&-i\dot{\phi}(t)\sigma'(t)+i\frac{\Omega(t)}{2}\Delta'(t)+iG\Delta'(t)\sigma'(t), \label{sigma_prime}\quad\quad
\end{eqnarray}
\end{subequations}
where observe that the difference with system (\ref{system1}) is that $\dot{\phi}$ is replaced by $-\dot{\phi}$. Now let us consider the transformation
\begin{subequations}
\label{transform}
\begin{eqnarray}
\Delta''(t)&=&\Delta'(t),\label{ddprime}\\
\sigma''(t)&=&-\sigma'^*(t)e^{2i\beta},\label{sdprime}
\end{eqnarray}
\end{subequations}
where recall that $\beta$ is the constant phase associated with the complex Rabi frequency, $\Omega(t)=|\Omega(t)|e^{i\beta}$.
It is not hard to show that the transformed variables satisfy the equations
\begin{subequations}
\label{system_dprime}
\begin{eqnarray}
\dot{\Delta}''(t)&=&i\Omega^*(t)\sigma''(t)-i\Omega(t)\sigma''^*(t)+4G_I\sigma''(t)\sigma''^*(t), \label{delta_dprime}\quad\quad\\
\dot{\sigma}''(t)&=&i\dot{\phi}(t)\sigma''(t)+i\frac{\Omega(t)}{2}\Delta''(t)-iG^*\Delta''(t)\sigma''(t), \label{sigma_dprime}\quad\quad
\end{eqnarray}
\end{subequations}
Observe that in Eqs. (\ref{system_dprime}) the chirp sign has been restored ($+\dot{\phi}$), and the only difference with Eqs. (\ref{system1}) is that $G$ is replaced by $-G^*=-G_R+iG_I$. For $G_R=0$ Eqs. (\ref{system1}) and (\ref{system_dprime}) are identical and, since the transformed variables satisfy also the same initial conditions $\Delta''(0)=1, \sigma''(0)=0$, we obtain that $\Delta''(t)=\Delta(t)$. But it is also $\Delta'(t)=\Delta''(t)$, thus $\Delta'(t)=\Delta(t)$ and the solutions corresponding to opposite chirp are equivalent. For $G_R\neq 0$ this symmetry breaks down. Note also that mathematically the symmetry is preserved if $G$ is replaced by $G'=-G^*$.

There is actually a simple intuitive explanation why the presence of $G_R$ breaks the ``$z$-symmetry" of the two-level system (Bloch sphere), which is revealed when we rewrite Eq. (\ref{sigma1}) for the coherence as
\begin{equation}
\label{coherence}
\dot{\sigma}(t)=i[\dot{\phi}(t)+G_R\Delta(t)]\sigma(t)+i\frac{\Omega(t)}{2}\Delta(t)-G_I\Delta(t)\sigma(t),
\end{equation}
using that $G=G_R+iG_I$. Now it is obvious that $G_R\neq 0$ results in an extra time-dependent field $G_R\Delta(t)$ in the ``$z$-direction", defined with respect to the two-level system (\ref{two_level}), which breaks the corresponding symmetry for positive and negative chirp $\dot{\phi}(t)$. In the two-level Schrodinger equation (\ref{two_level}) the $\dot{\phi}(t)$ term should be replaced by $\dot{\phi}(t)+G_R\Delta(t)$ and the instantaneous ``eigenvalues" (\ref{eigenvalues}) become state-dependent
\begin{equation}\label{modified_eigenvalues}
A_{\pm}(t)=\pm\frac{1}{2}\sqrt{(\dot{\phi}(t)+G_R\Delta(t))^2+|\Omega(t)|^2} \, ,
\end{equation}
where for the latter term we use quotation marks since in the presence of the MNP the system is nonlinear. In the next section we present specific examples of how the added term $G_R\Delta(t)$ modifies the gap between the ``eigenvalues" for the different chirp signs.
Note that in Eq. (\ref{two_level}), the term $\dot{\phi}(t)$ multiplies the $-\sigma_z$ Pauli spin matrix while the term $\Omega(t)$ multiplies a linear combination of the $\sigma_x, \sigma_y$ matrices, and this is why we characterize them as the ``longitudinal" and ``transverse" fields, respectively. Nevertheless, we point out that the ``transverse" field $\Omega(t)$ for the two-level system corresponds to the applied field, which points in the $z$-direction in the real space.

\section{Numerical simulations and results analysis}

\label{sec:results}

\begin{figure*}[t]
 \centering
		\begin{tabular}{cc}
        \subfigure[$\ $]{
	            \label{fig:G_1_11}
	            \includegraphics[width=.45\linewidth]{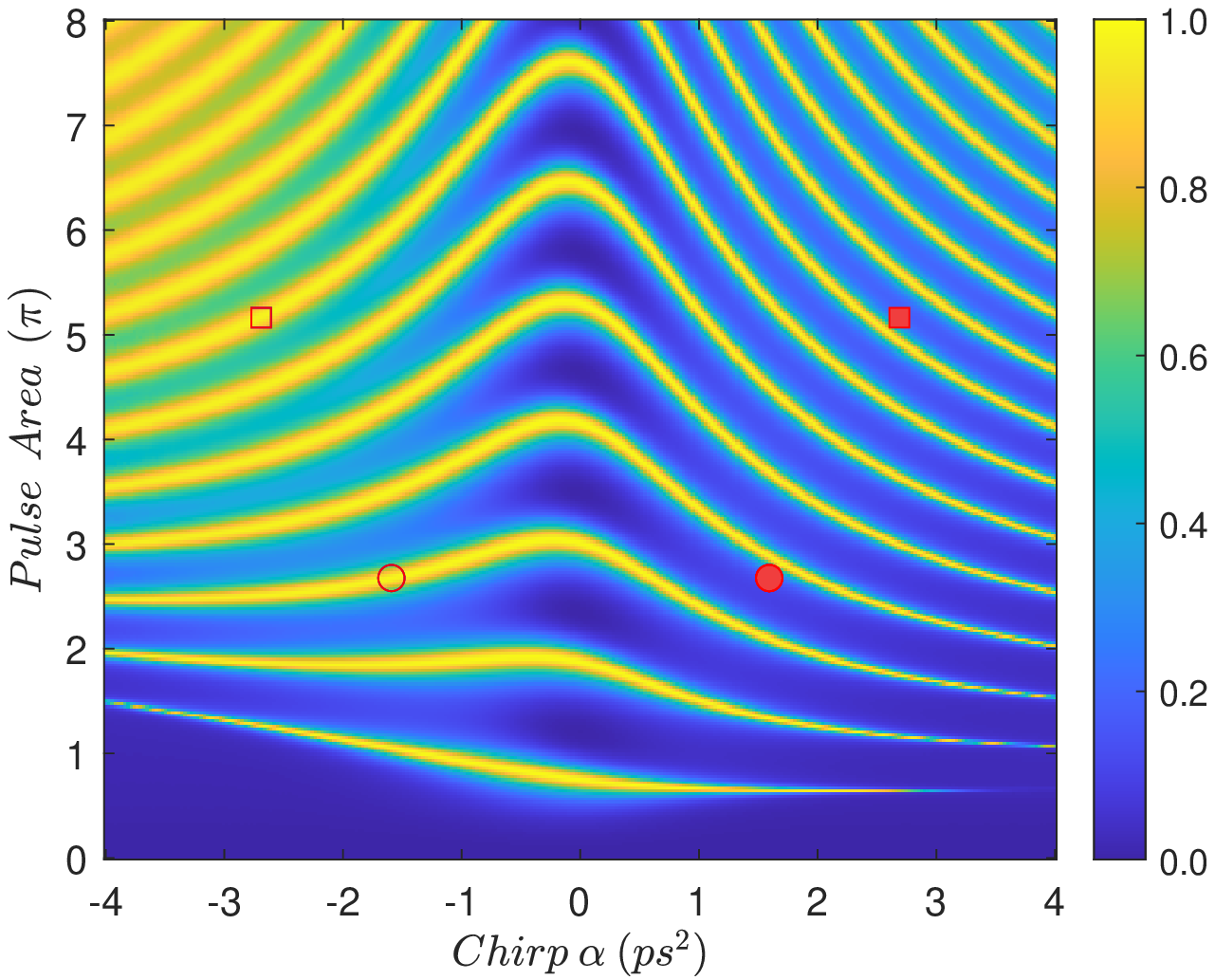}} &
        \subfigure[$\ $]{
	            \label{fig:G_1_12}
	            \includegraphics[width=.45\linewidth]{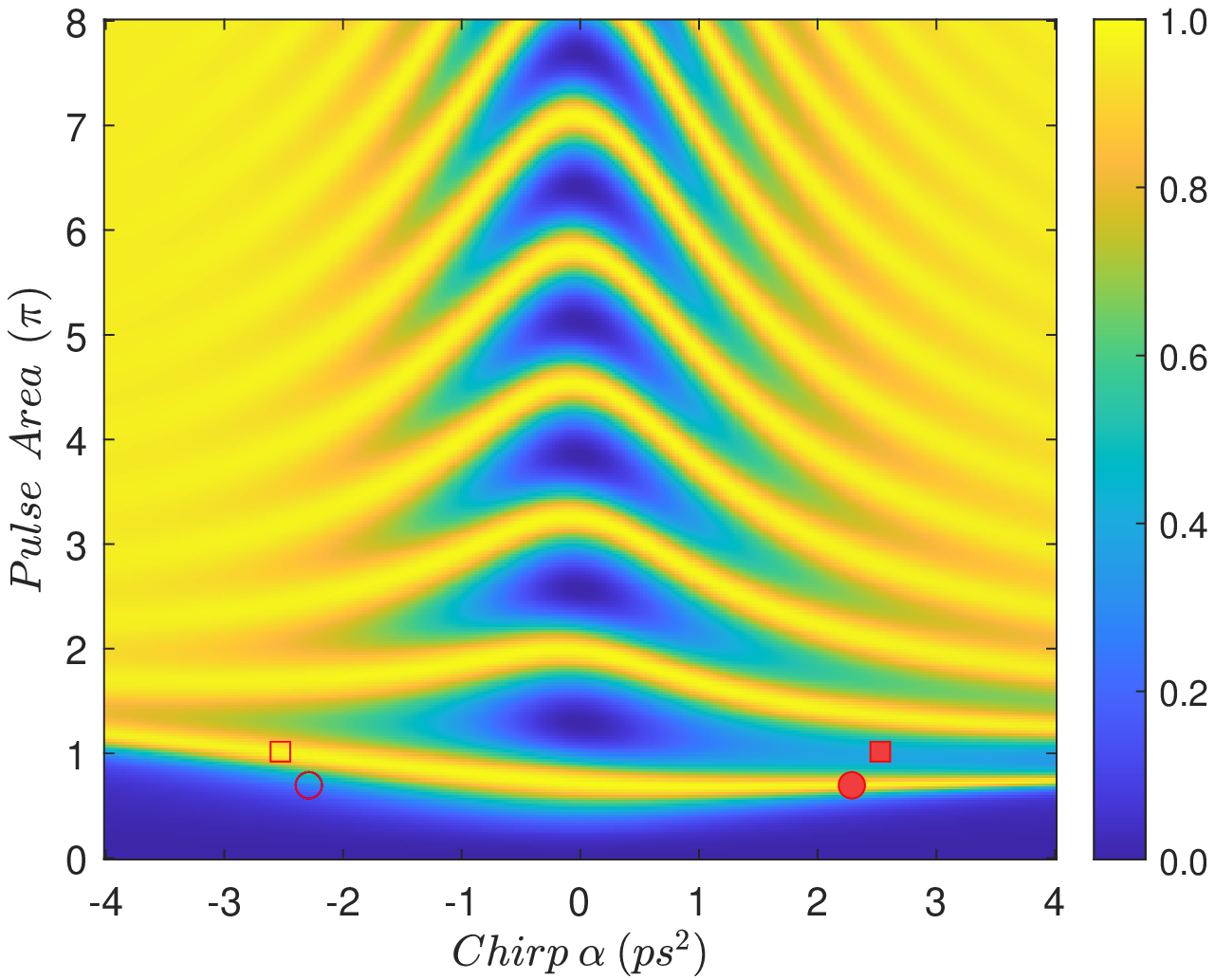}} \\
        \subfigure[$\ $]{
	            \label{fig:G_1_13}
	            \includegraphics[width=.45\linewidth]{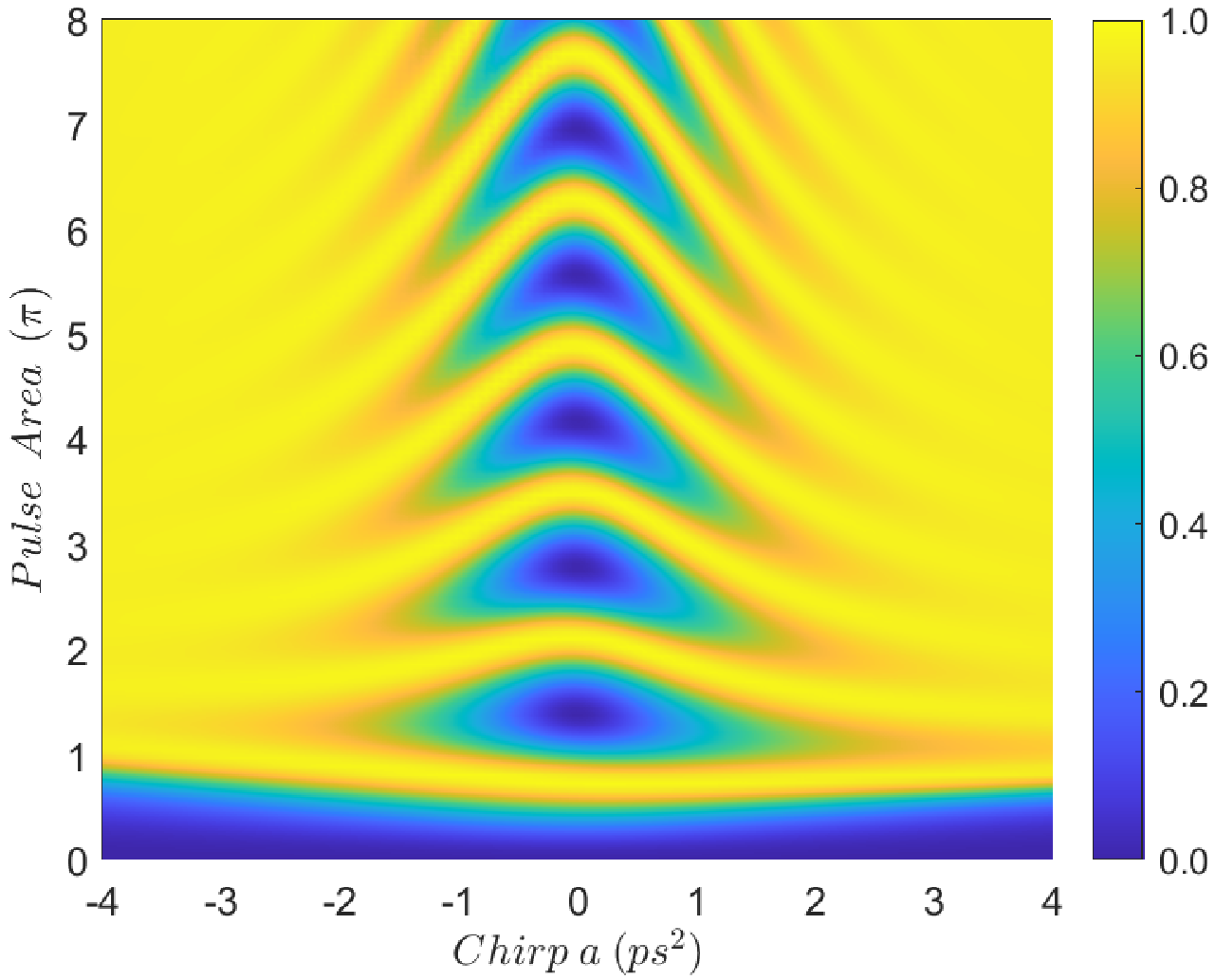}} &
        \subfigure[$\ $]{
	            \label{fig:G_1_15}
	            \includegraphics[width=.45\linewidth]{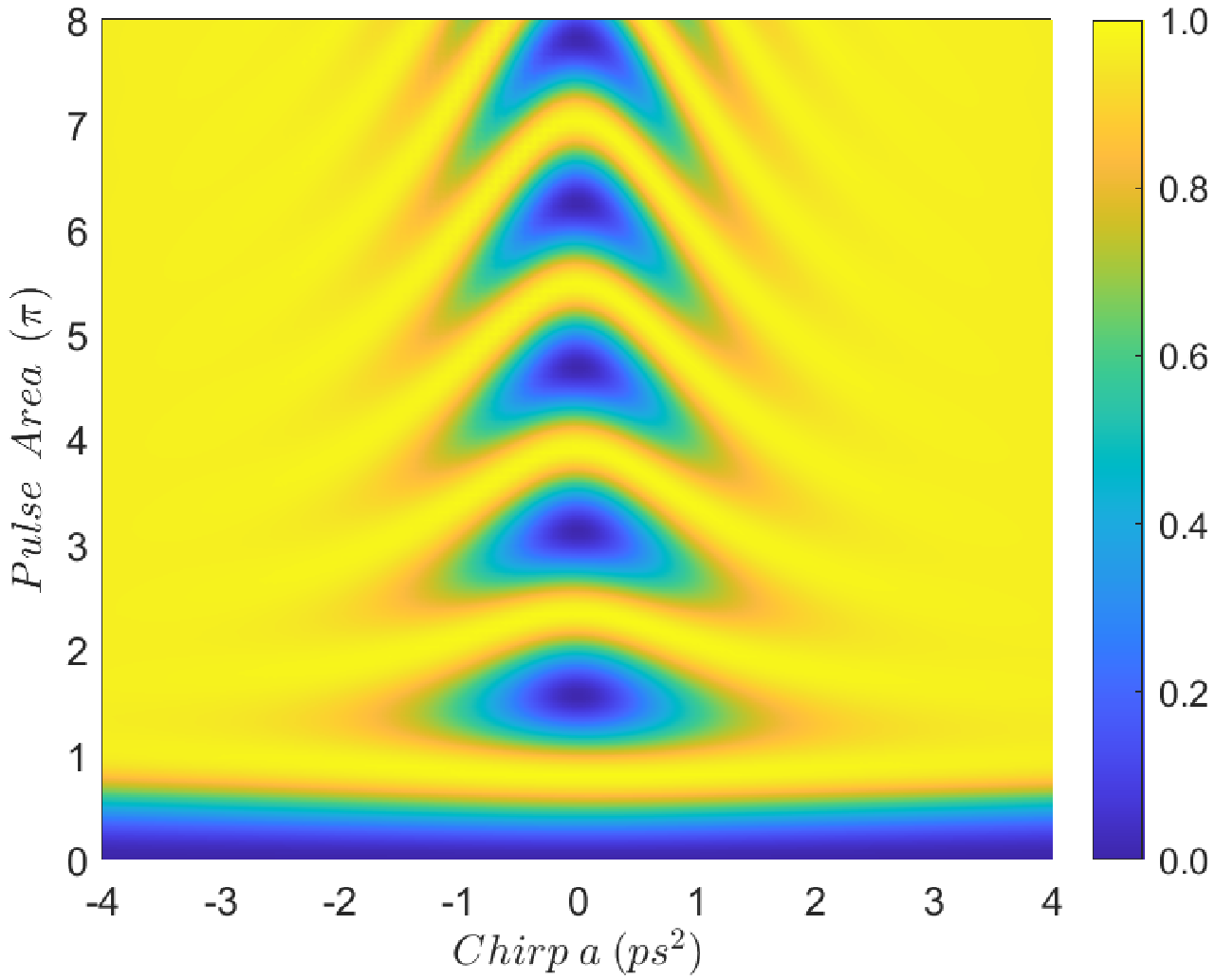}} \\
        \subfigure[$\ $]{
	            \label{fig:G_1_30}
	            \includegraphics[width=.45\linewidth]{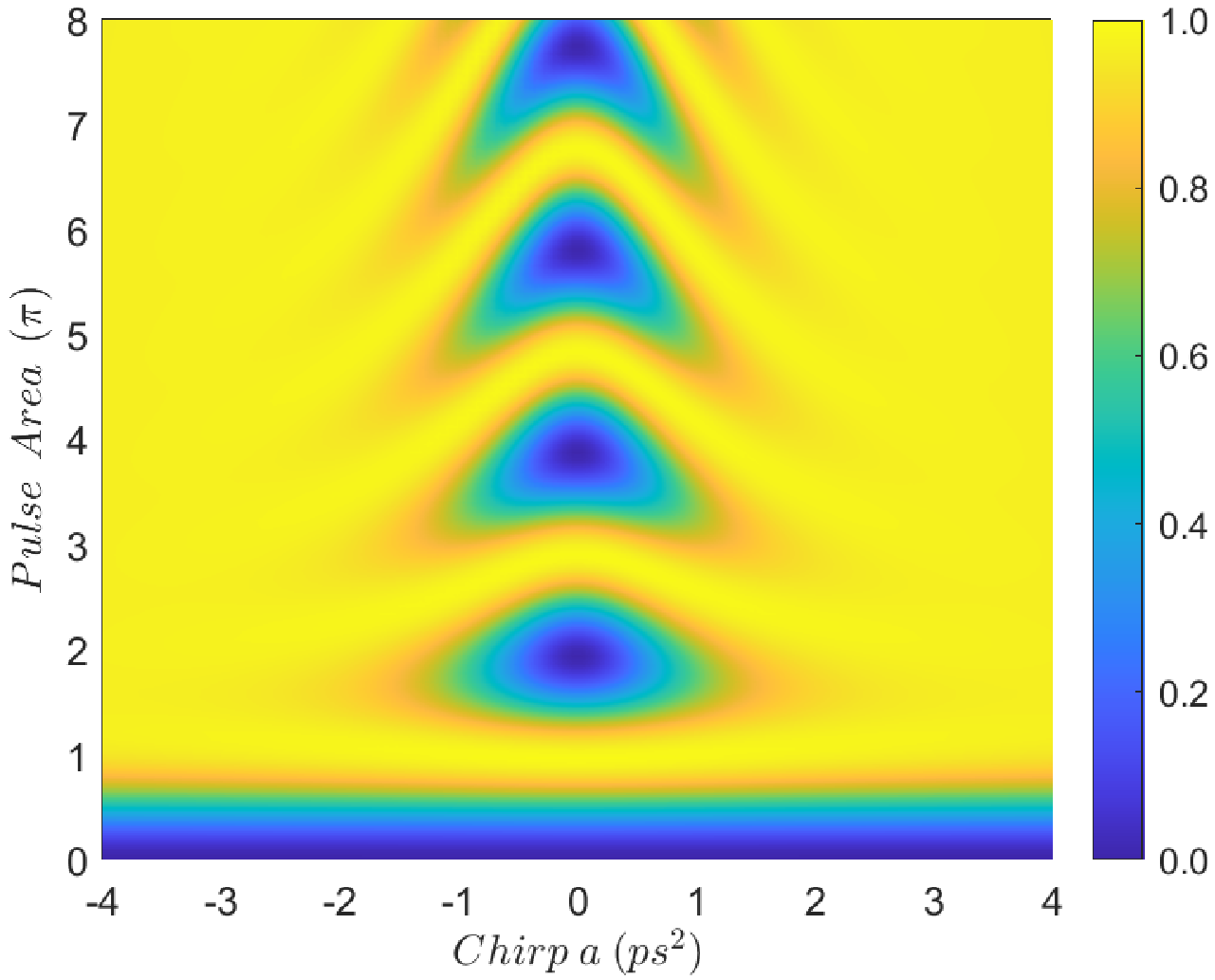}} &
        \subfigure[$\ $]{
	            \label{fig:G_1_80}
	            \includegraphics[width=.45\linewidth]{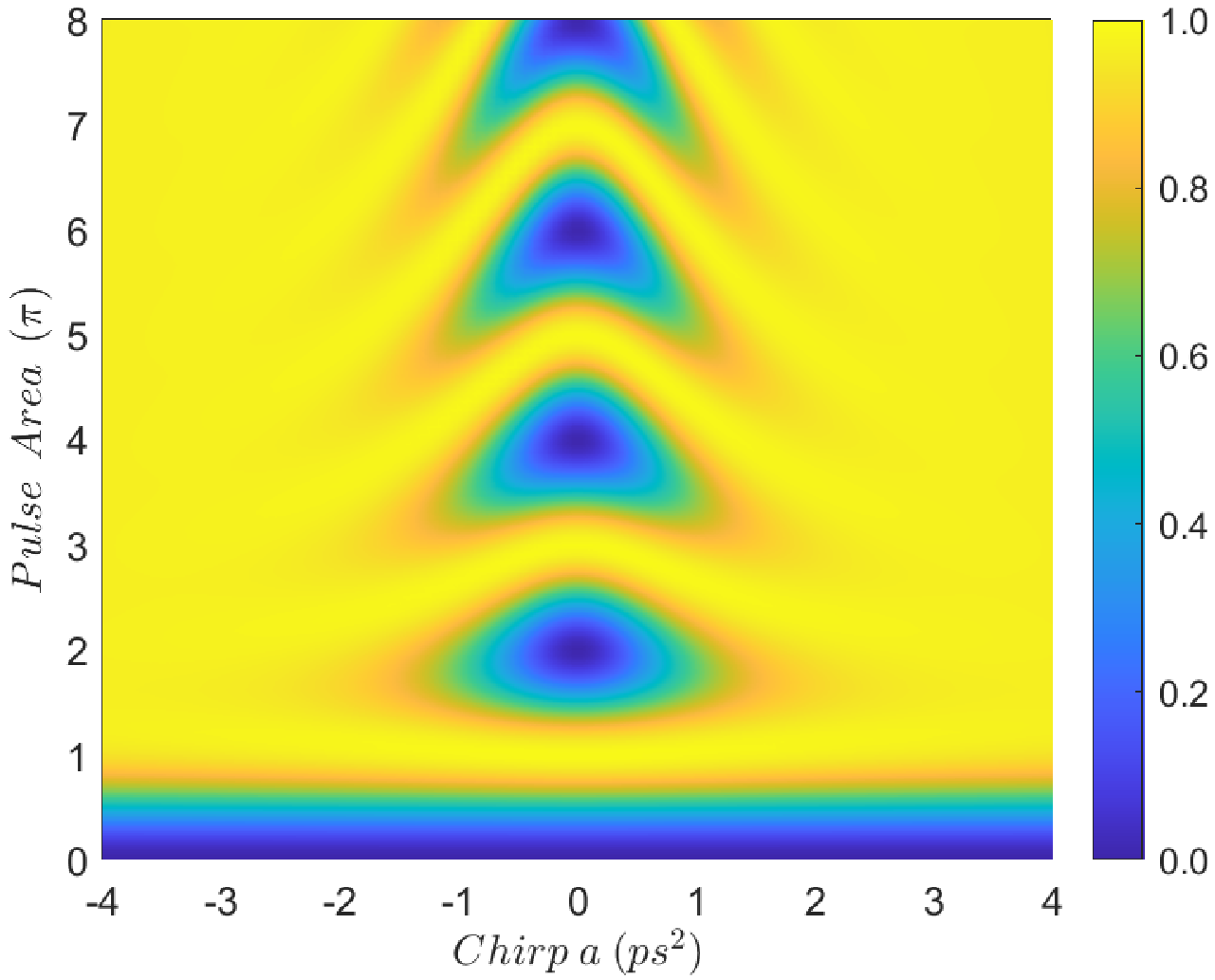}}
		\end{tabular}
\caption{(Color online) Contour plot of the final exciton population when using the Gaussian chirped pulse with $\tau_0=1$ ps, versus the pulse area and the chirp parameter $a$, for different values of the interparticle distance: (a) $R=11$ nm, (b) $R=12$ nm, (c) $R=13$ nm, (d) $R=15$ nm, (e) $R=30$ nm, (f) $R=80$ nm.}
\label{fig:G_1}
\end{figure*}

\begin{figure*}[t]
 \centering
		\begin{tabular}{cc}
        \subfigure[$\ $]{
	            \label{fig:G_075_11}
	            \includegraphics[width=.45\linewidth]{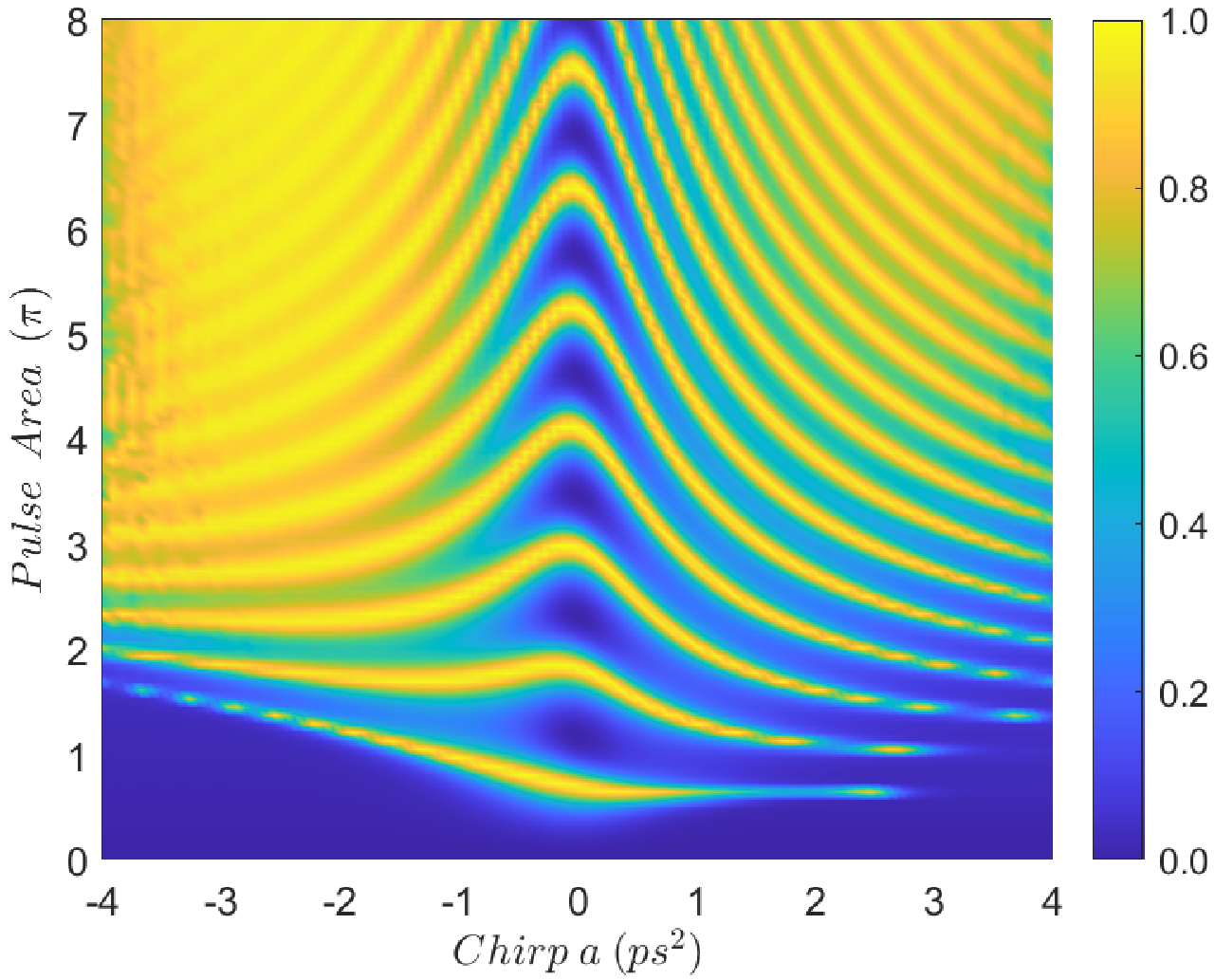}} &
        \subfigure[$\ $]{
	            \label{fig:G_075_12}
	            \includegraphics[width=.45\linewidth]{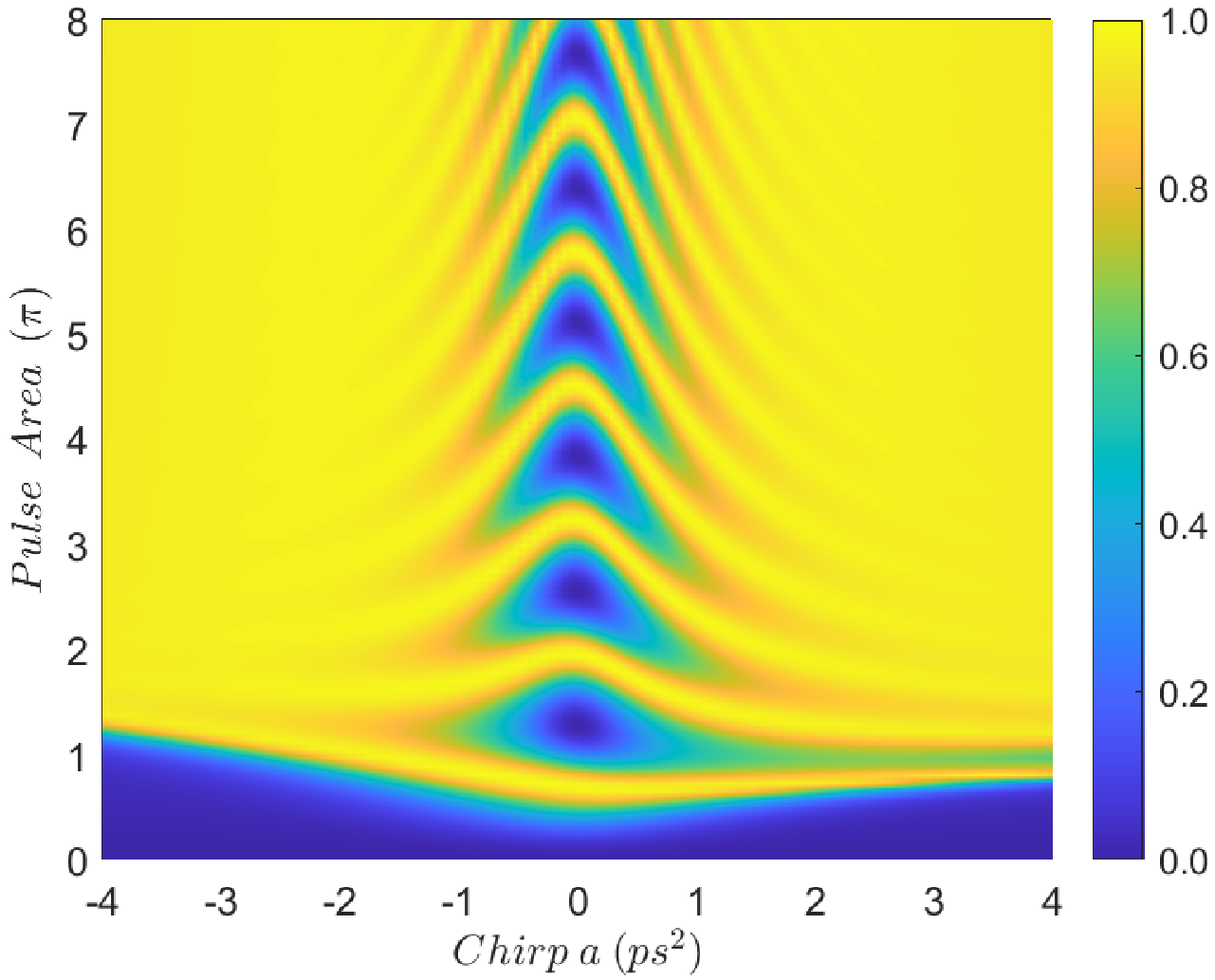}} \\
        \subfigure[$\ $]{
	            \label{fig:G_075_13}
	            \includegraphics[width=.45\linewidth]{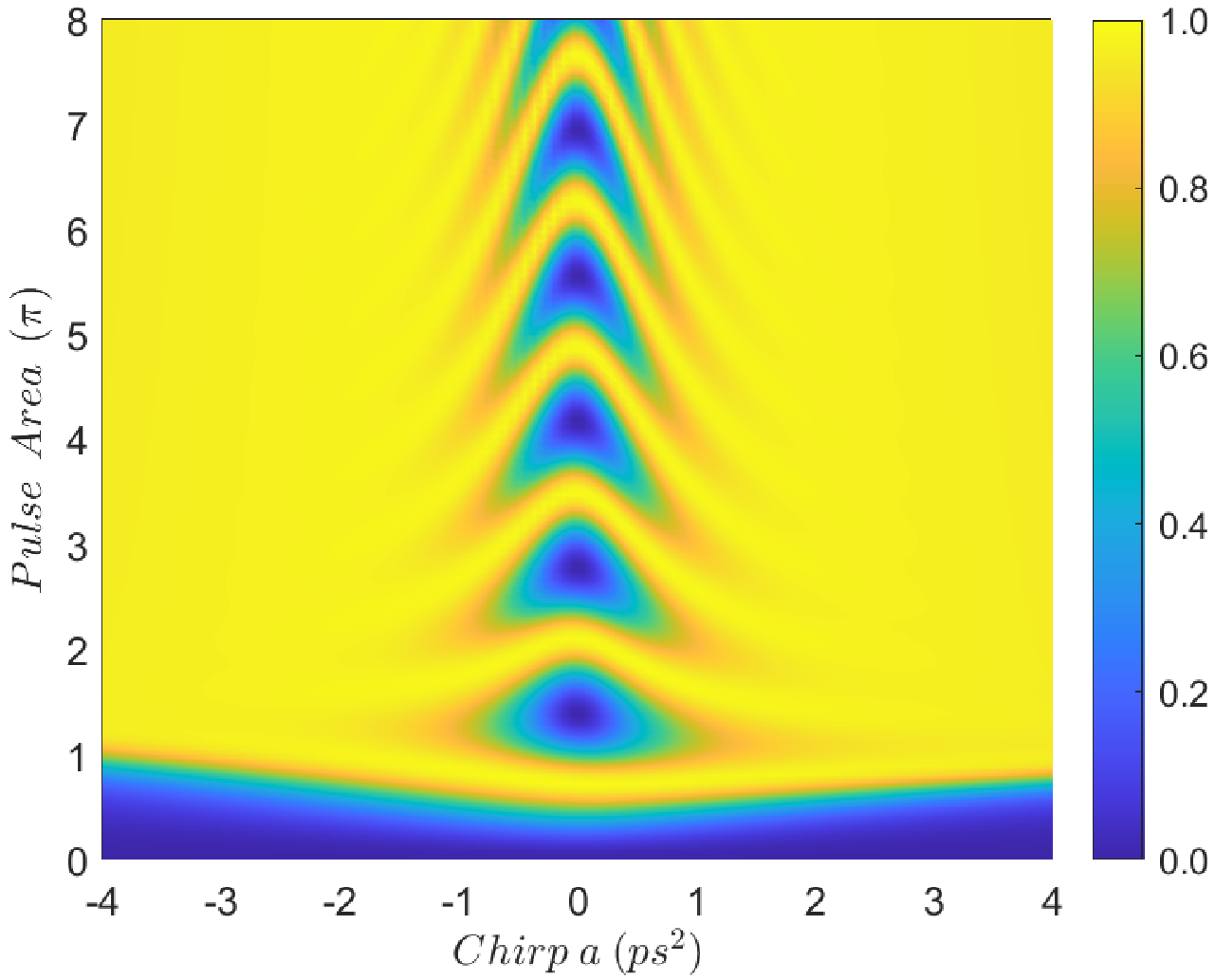}} &
        \subfigure[$\ $]{
	            \label{fig:G_075_15}
	            \includegraphics[width=.45\linewidth]{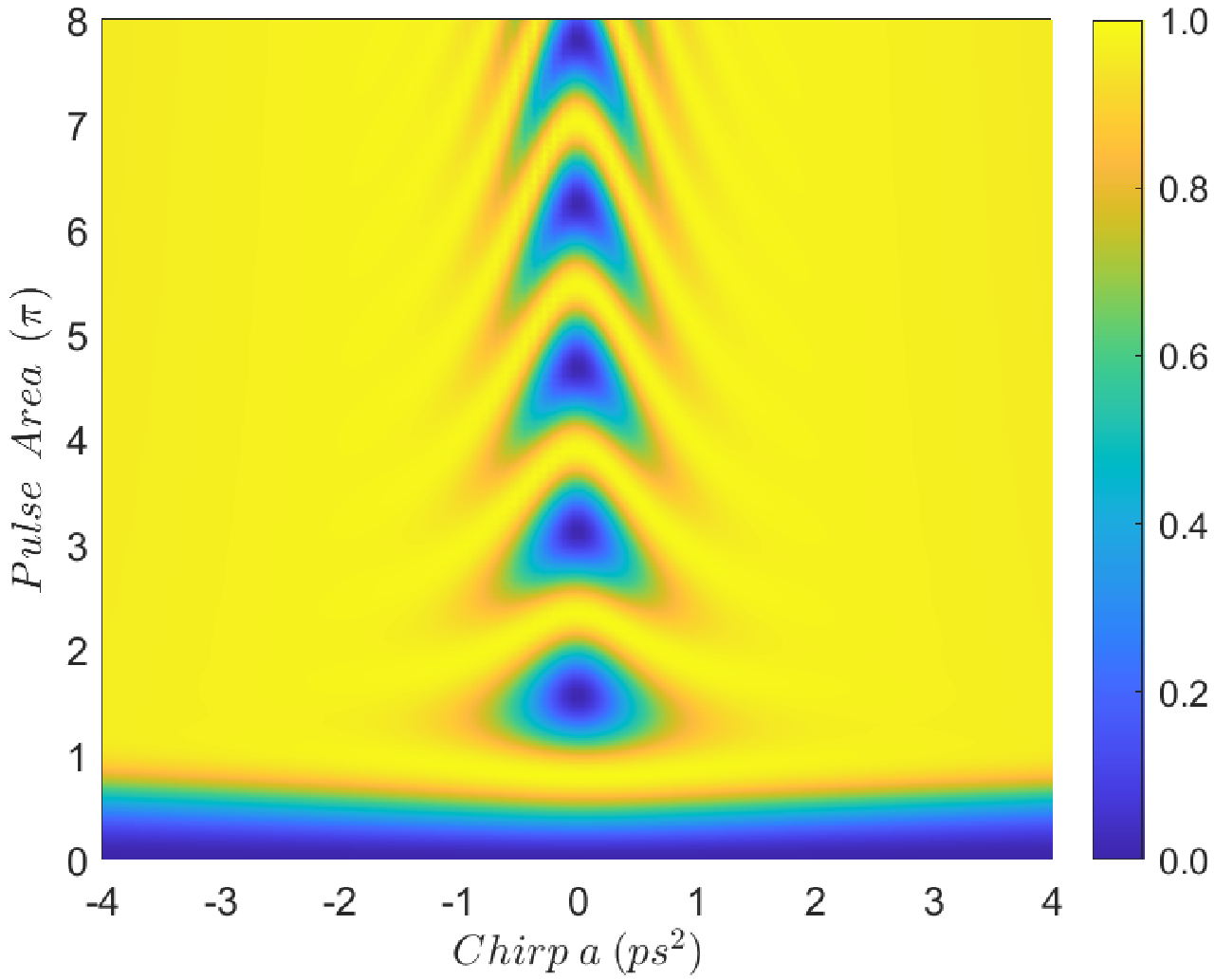}} \\
        \subfigure[$\ $]{
	            \label{fig:G_075_30}
	            \includegraphics[width=.45\linewidth]{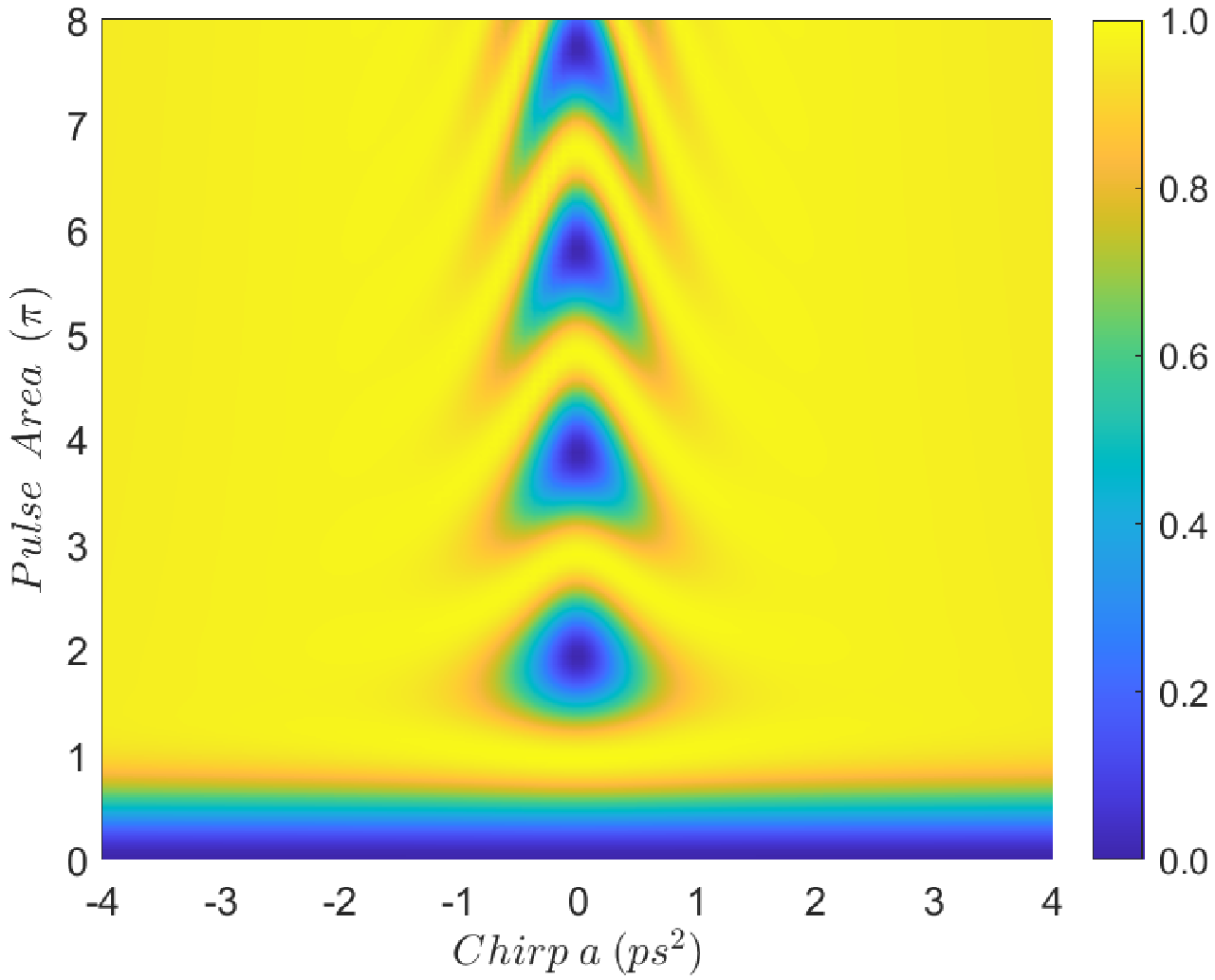}} &
        \subfigure[$\ $]{
	            \label{fig:G_075_80}
	            \includegraphics[width=.45\linewidth]{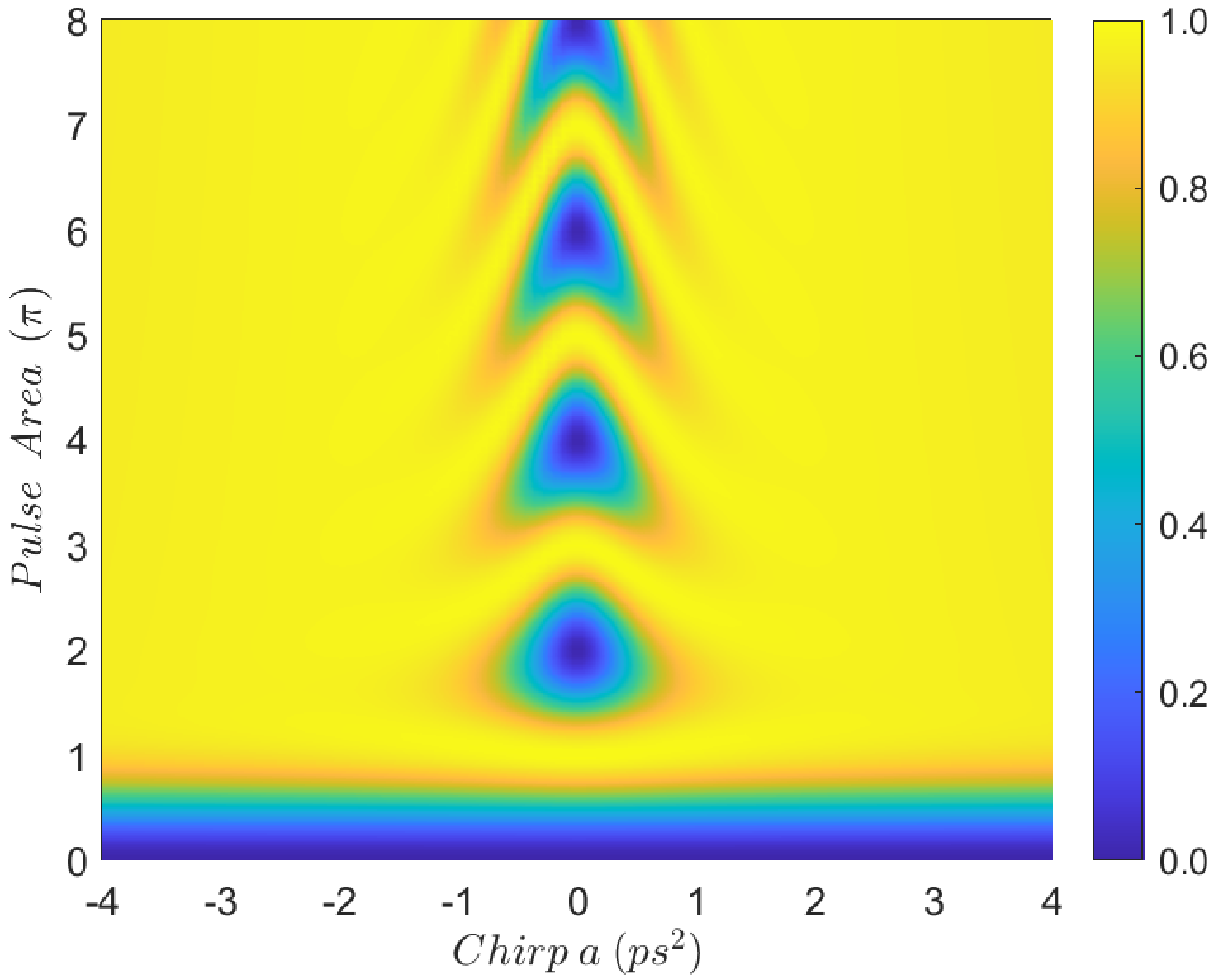}}
		\end{tabular}
\caption{(Color online) Contour plot of the final exciton population when using the Gaussian chirped pulse with $\tau_0=0.75$ ps, versus the pulse area and the chirp parameter $a$, for different values of the interparticle distance: (a) $R=11$ nm, (b) $R=12$ nm, (c) $R=13$ nm, (d) $R=15$ nm, (e) $R=30$ nm, (f) $R=80$ nm.}
\label{fig:G_075}
\end{figure*}

\begin{figure}[t]
 \centering
		\begin{tabular}{c}
        \subfigure[$\ $]{
	            \label{fig:GR_1}
	            \includegraphics[width=0.85\linewidth]{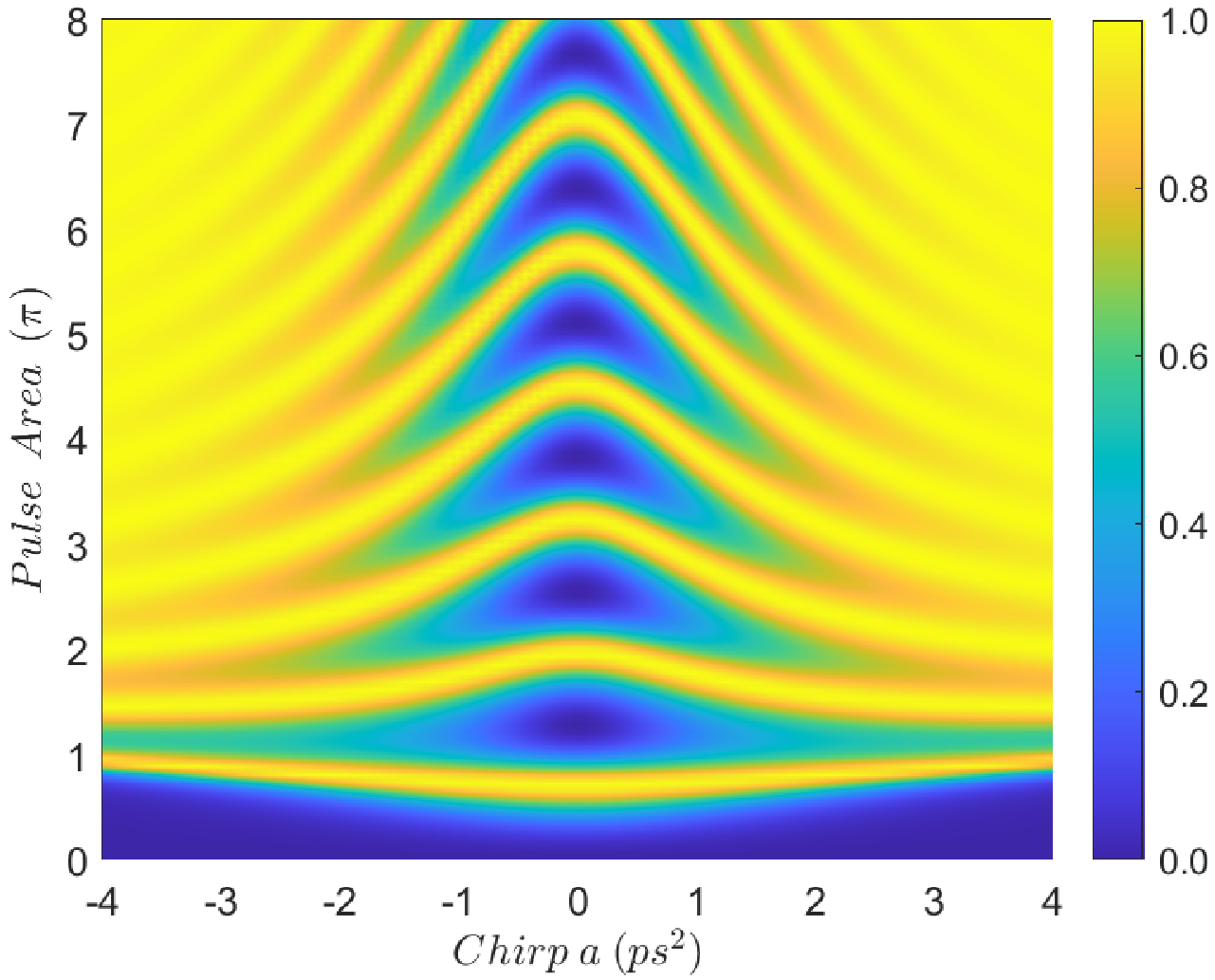}} \\
        \subfigure[$\ $]{
	            \label{fig:GR_075}
	            \includegraphics[width=0.85\linewidth]{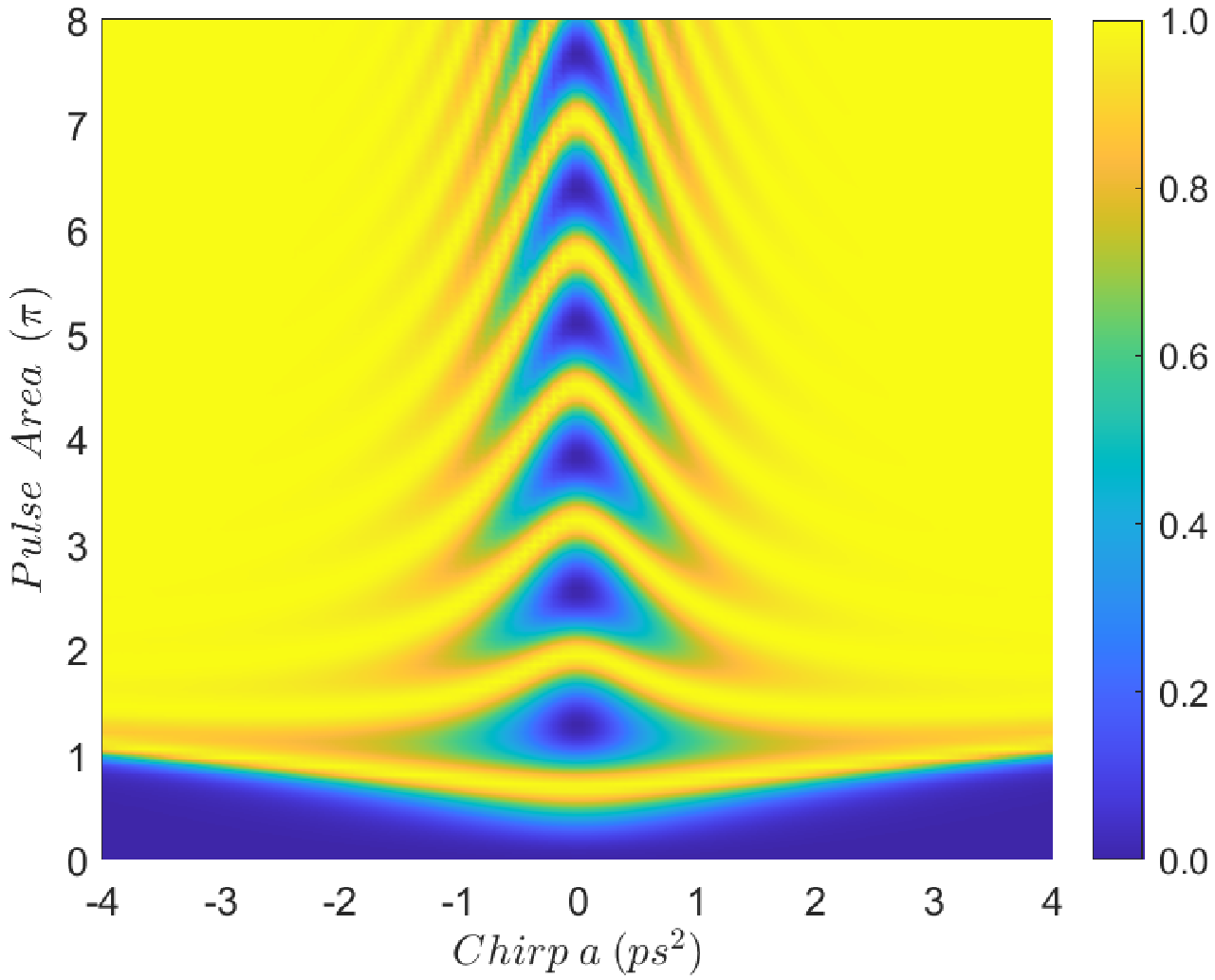}}
		\end{tabular}
\caption{(Color online) Contour plot of the final exciton population versus the pulse area and the chirp parameter $a$ when manually enforcing $G_R=\mbox{Re}\{G\}=0$, for $R=12$ nm and Gaussian chirped pulse with (a) $\tau_0=1$ ps, (b) $\tau_0=0.75$ ps.}
\label{fig:GR}
\end{figure}

\begin{figure}[t]
 \centering
		\begin{tabular}{c}
        \subfigure[$\ $]{
	            \label{fig:m_GR_1}
	            \includegraphics[width=0.85\linewidth]{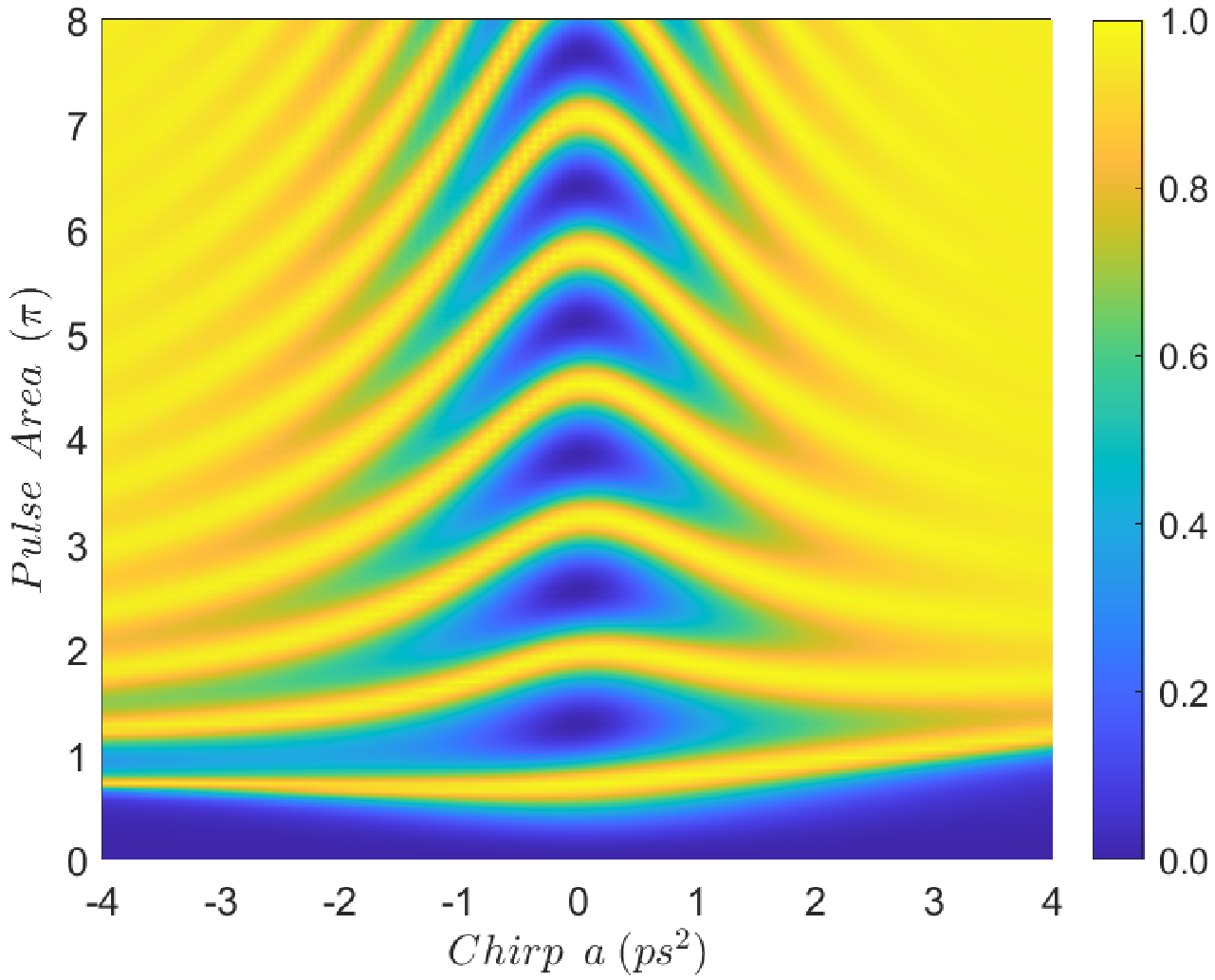}} \\
        \subfigure[$\ $]{
	            \label{fig:m_GR_075}
	            \includegraphics[width=0.85\linewidth]{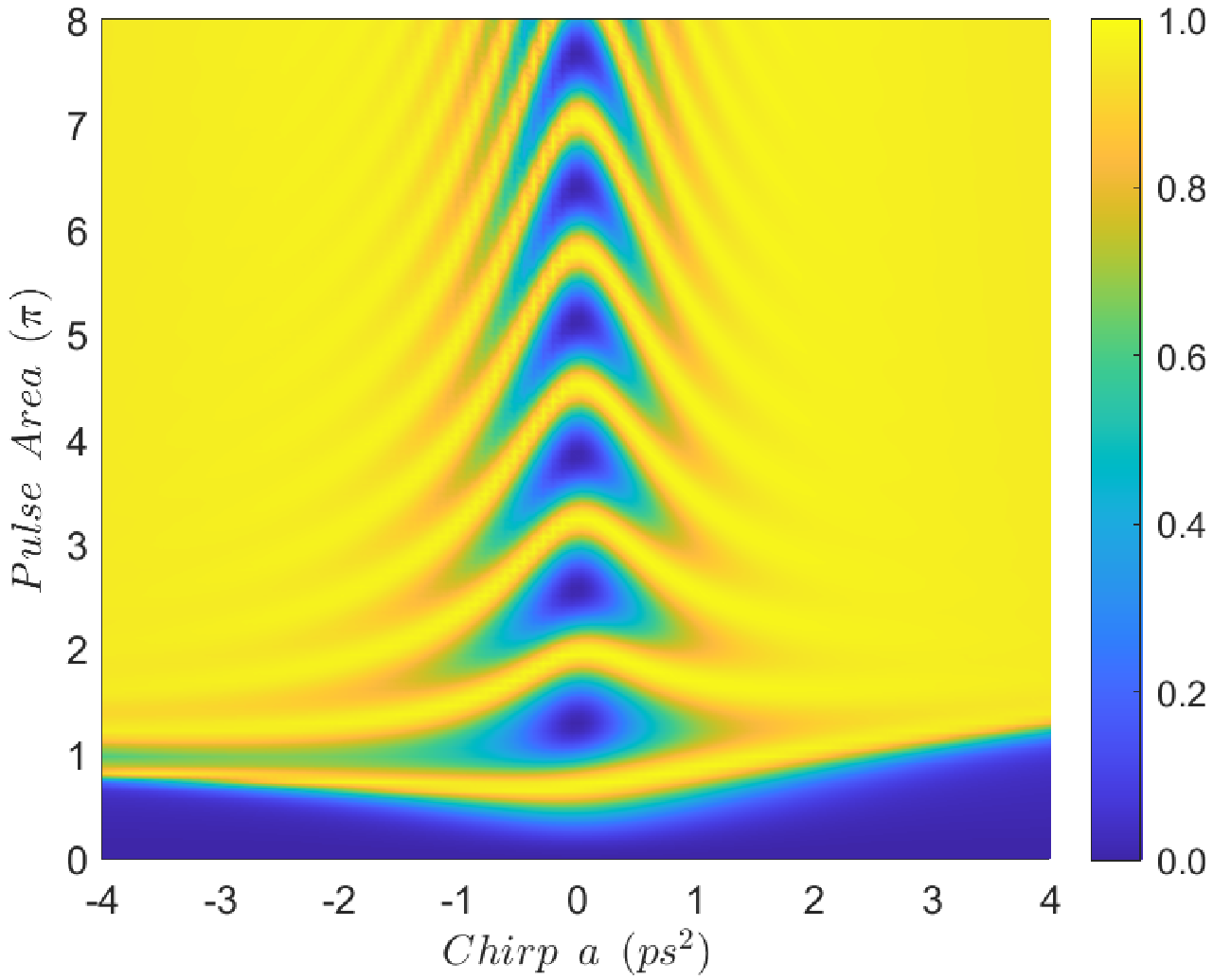}}
		\end{tabular}
\caption{(Color online) Contour plot of the final exciton population versus the pulse area and the chirp parameter $a$ when manually inverting the sign of $G_R=\mbox{Re}\{G\}$, for $R=12$ nm and Gaussian chirped pulse with (a) $\tau_0=1$ ps, (b) $\tau_0=0.75$ ps.}
\label{fig:m_GR}
\end{figure}

\begin{figure*}[t]
 \centering
		\begin{tabular}{cc}
        \subfigure[$\ $]{
	            \label{fig:gap_11_1}
	            \includegraphics[width=.45\linewidth]{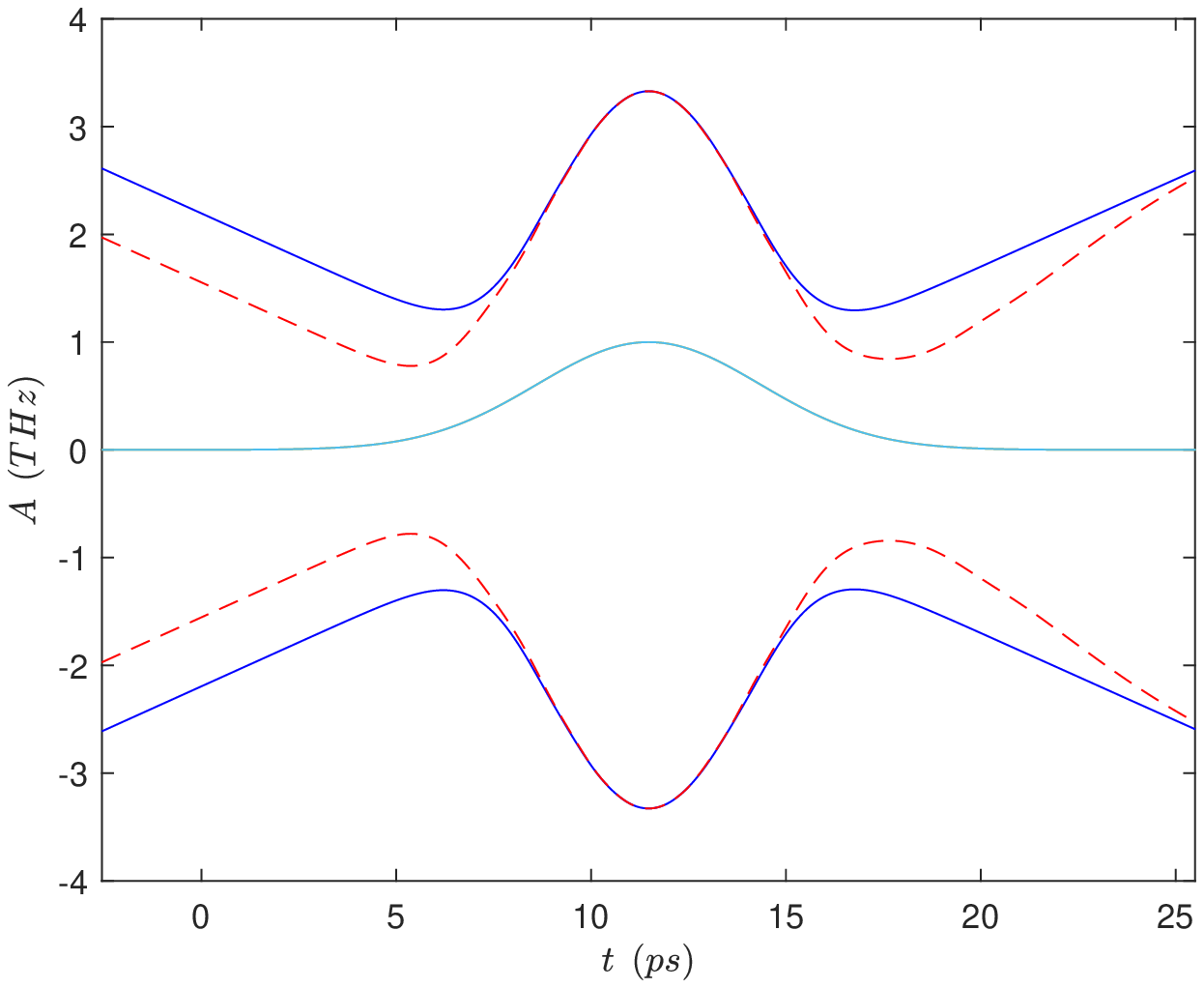}} &
        \subfigure[$\ $]{
	            \label{fig:gap_11_2}
	            \includegraphics[width=.45\linewidth]{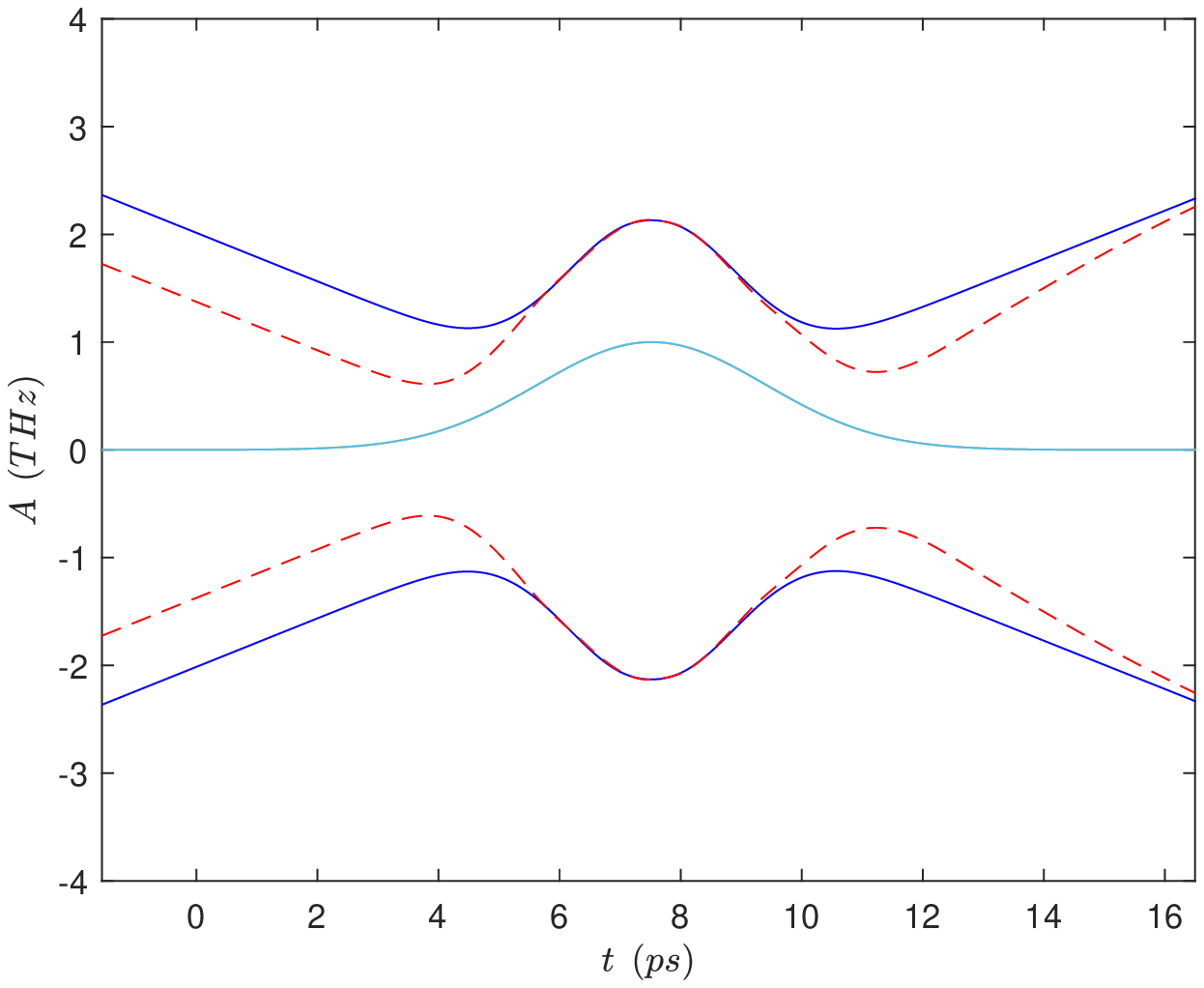}} \\
        \subfigure[$\ $]{
	            \label{fig:gap_12_1}
	            \includegraphics[width=.45\linewidth]{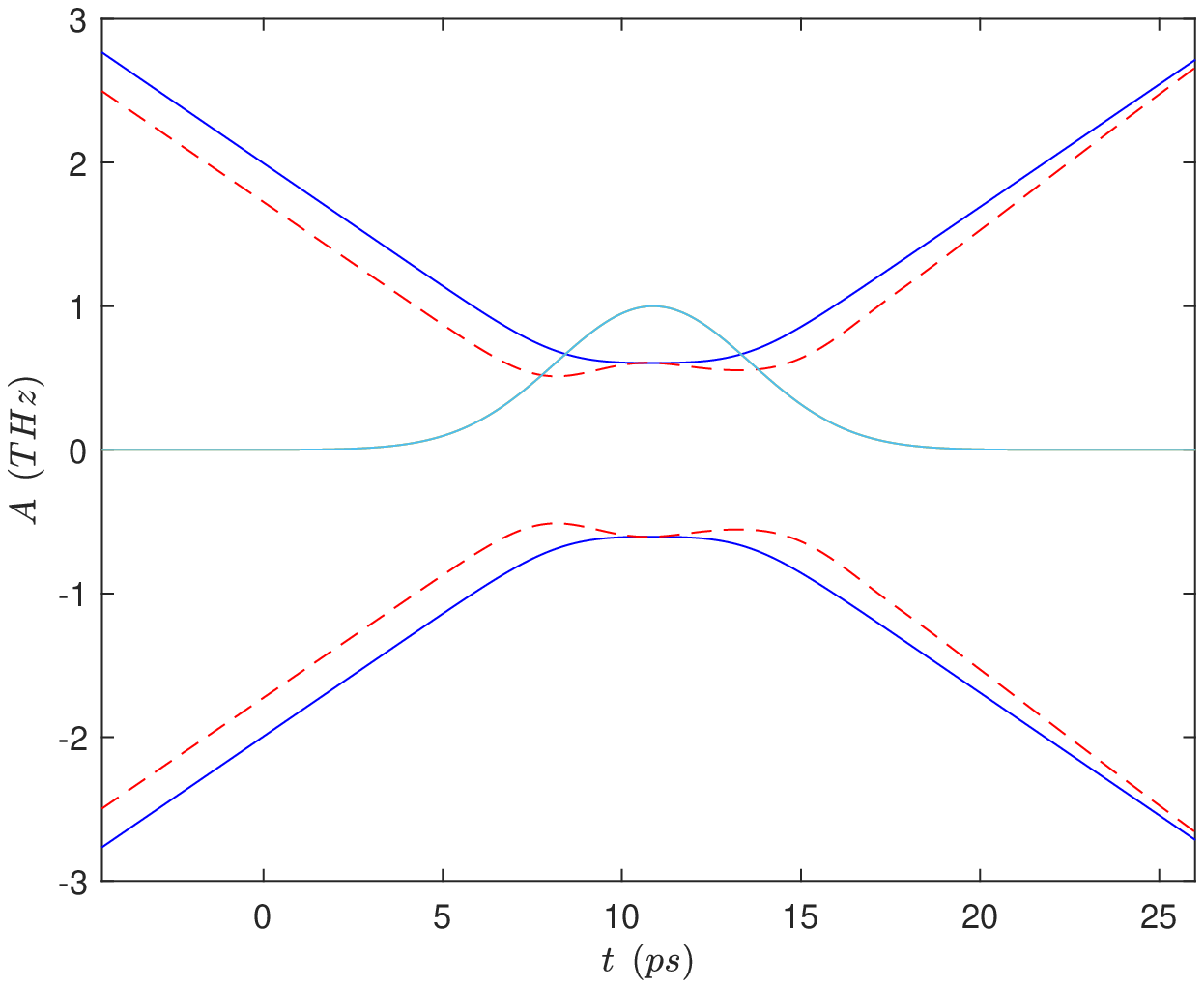}} &
        \subfigure[$\ $]{
	            \label{fig:gap_12_2}
	            \includegraphics[width=.45\linewidth]{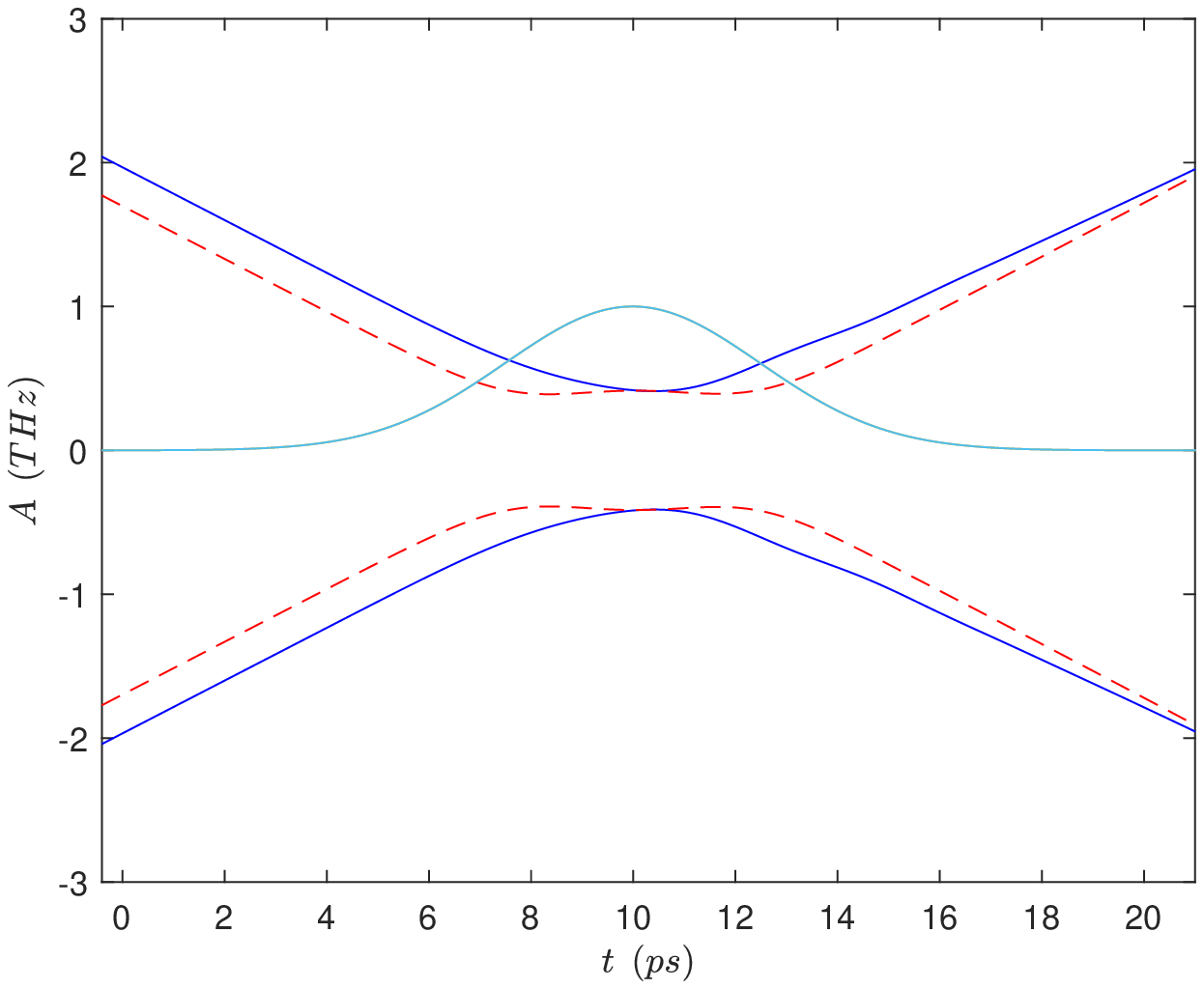}}
		\end{tabular}
\caption{(Color online) Modified ``eigenvalues" (\ref{modified_eigenvalues}) for opposite chirp values. Red dashed curves correspond to positive chirp while blue solid curves to the opposite negative chirp. The parameters used for each case correspond to: (a) upper symmetric pair marked in Fig. \ref{fig:G_1_11}, (b) lower symmetric pair marked in Fig. \ref{fig:G_1_11}, (c) upper symmetric pair marked in \ref{fig:G_1_12}, (d) lower symmetric pair marked in Fig. \ref{fig:G_1_12}.}
\label{fig:gap}
\end{figure*}

\begin{figure*}[t]
 \centering
		\begin{tabular}{cc}
        \subfigure[$\ $]{
	            \label{fig:S_15_11}
	            \includegraphics[width=.45\linewidth]{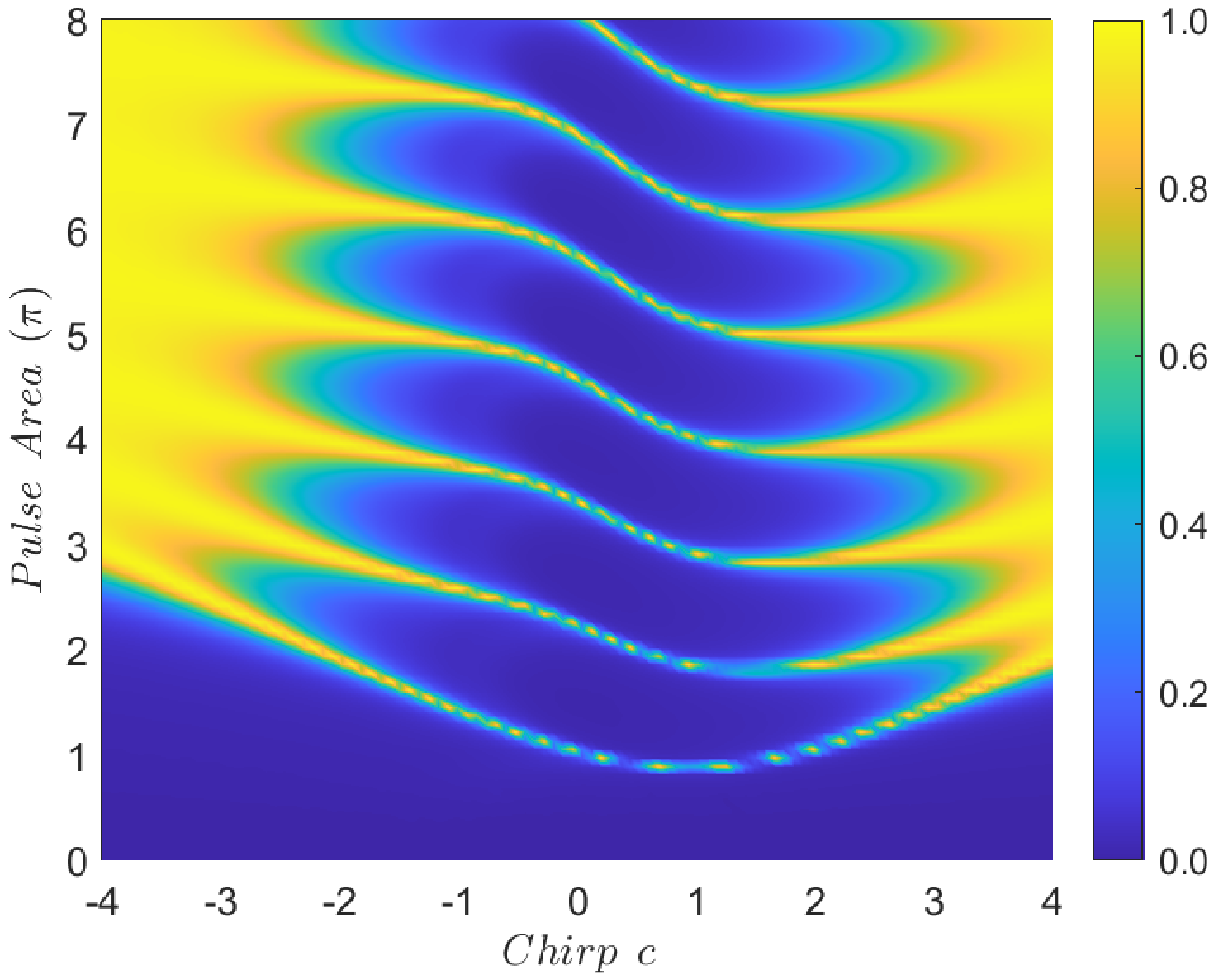}} &
        \subfigure[$\ $]{
	            \label{fig:S_15_12}
	            \includegraphics[width=.45\linewidth]{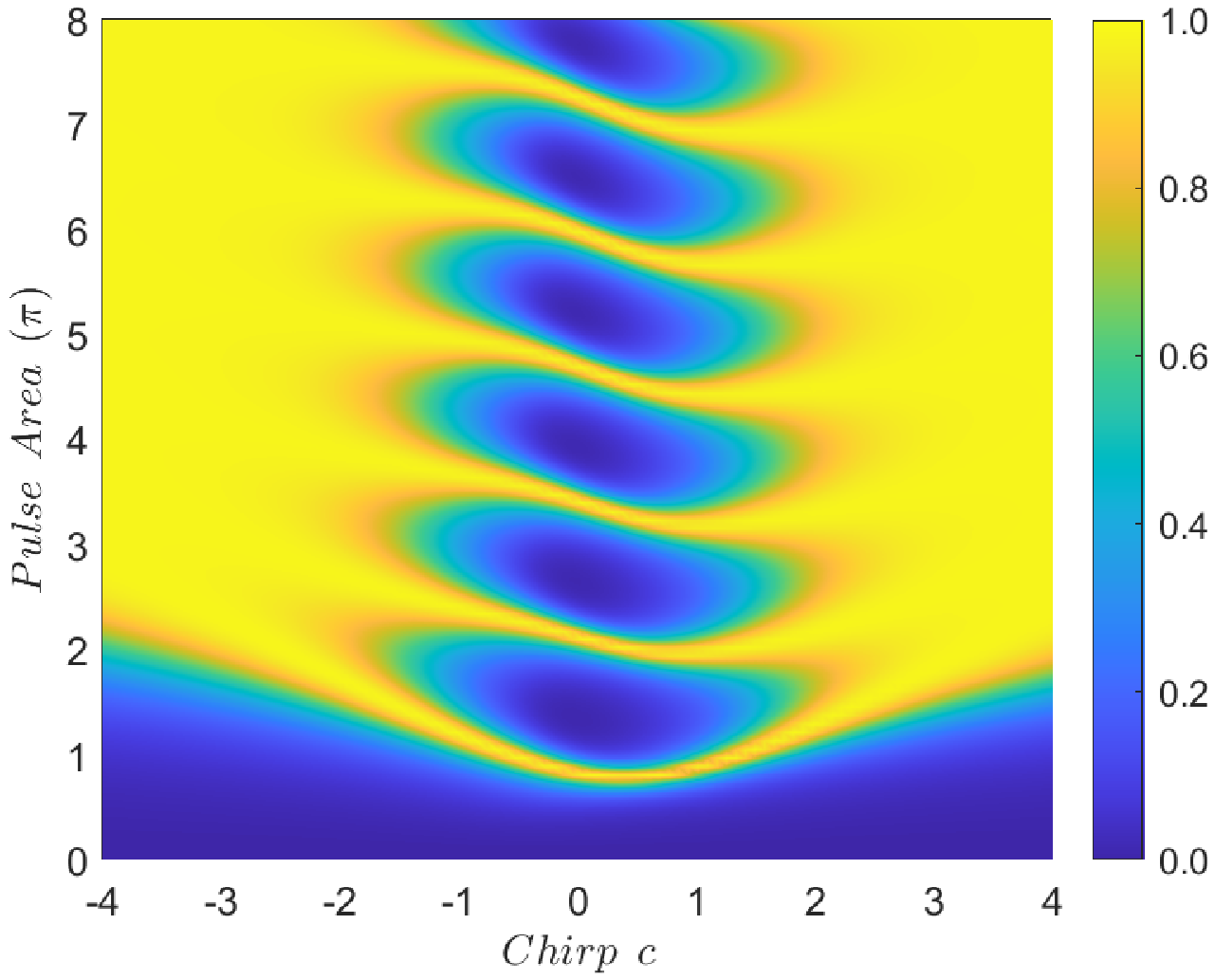}} \\
        \subfigure[$\ $]{
	            \label{fig:S_15_13}
	            \includegraphics[width=.45\linewidth]{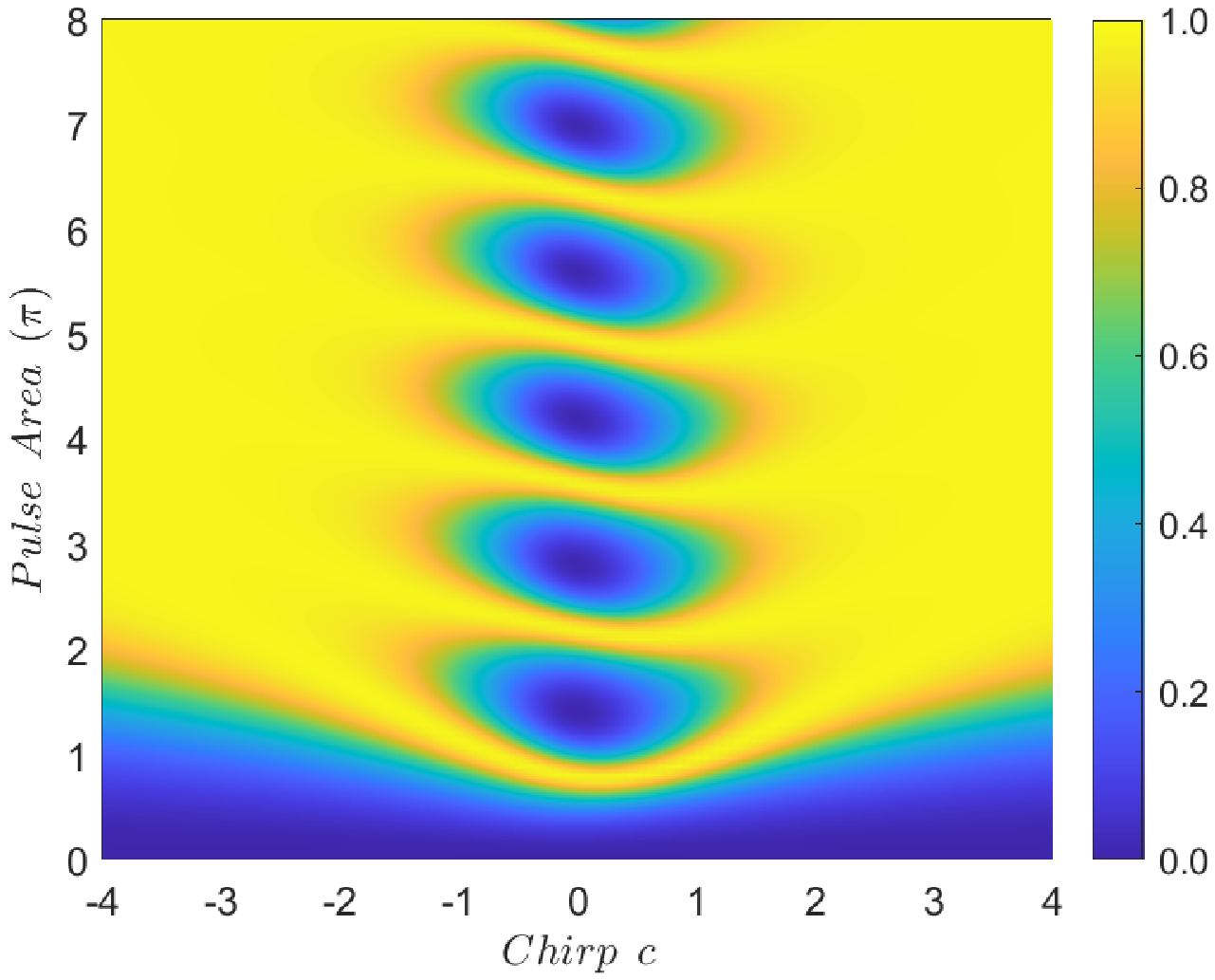}} &
        \subfigure[$\ $]{
	            \label{fig:S_15_15}
	            \includegraphics[width=.45\linewidth]{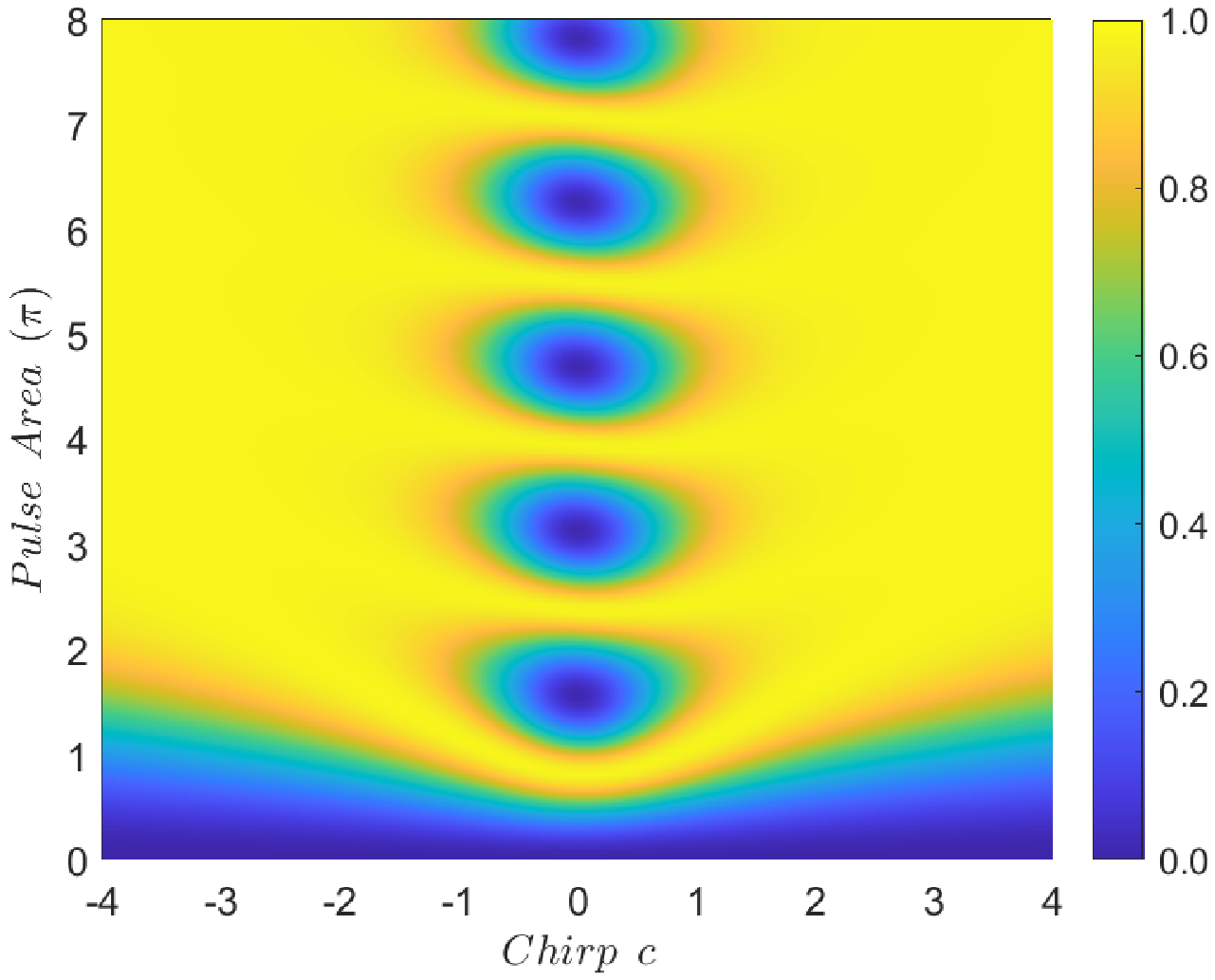}} \\
        \subfigure[$\ $]{
	            \label{fig:S_15_30}
	            \includegraphics[width=.45\linewidth]{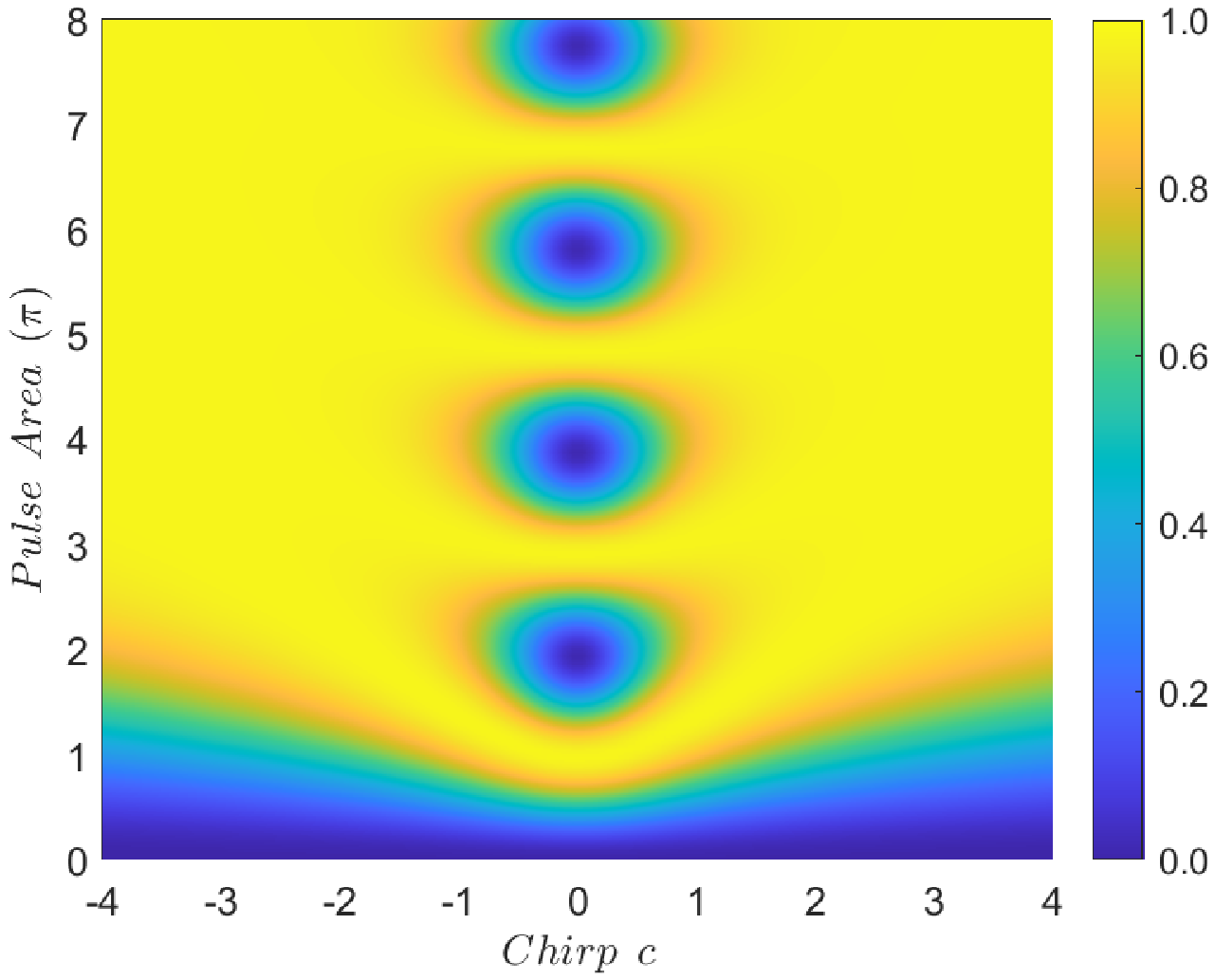}} &
        \subfigure[$\ $]{
	            \label{fig:S_15_80}
	            \includegraphics[width=.45\linewidth]{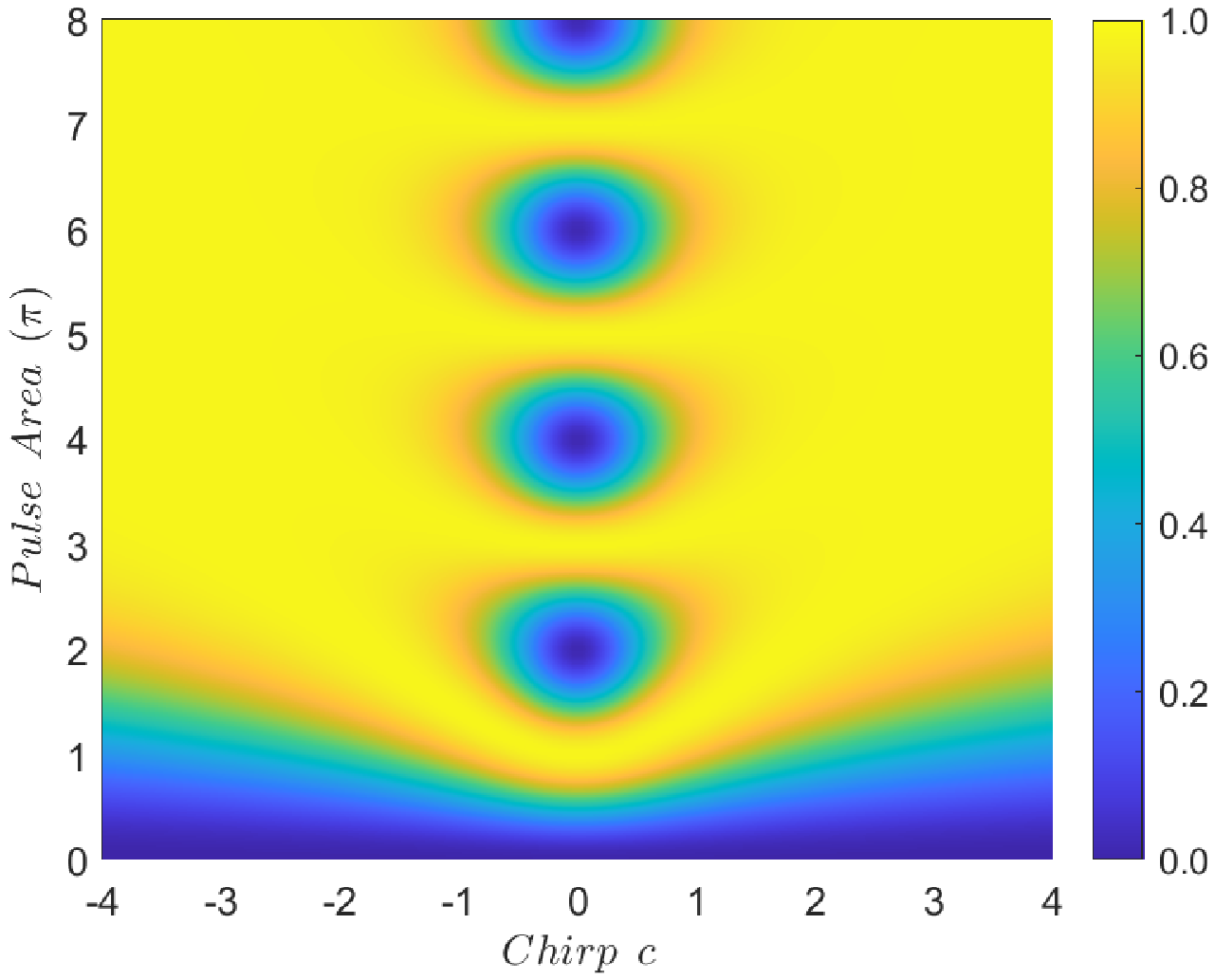}}
		\end{tabular}
\caption{(Color online) Contour plot of the final exciton population when using the hyperbolic secant chirped pulse with $t_p=1.5$ ps, versus the pulse area and the chirp parameter $c$, for different values of the interparticle distance: (a) $R=11$ nm, (b) $R=12$ nm, (c) $R=13$ nm, (d) $R=15$ nm, (e) $R=30$ nm, (f) $R=80$ nm.}
\label{fig:S_15}
\end{figure*}

\begin{figure*}[t]
 \centering
		\begin{tabular}{cc}
        \subfigure[$\ $]{
	            \label{fig:S_075_11}
	            \includegraphics[width=.45\linewidth]{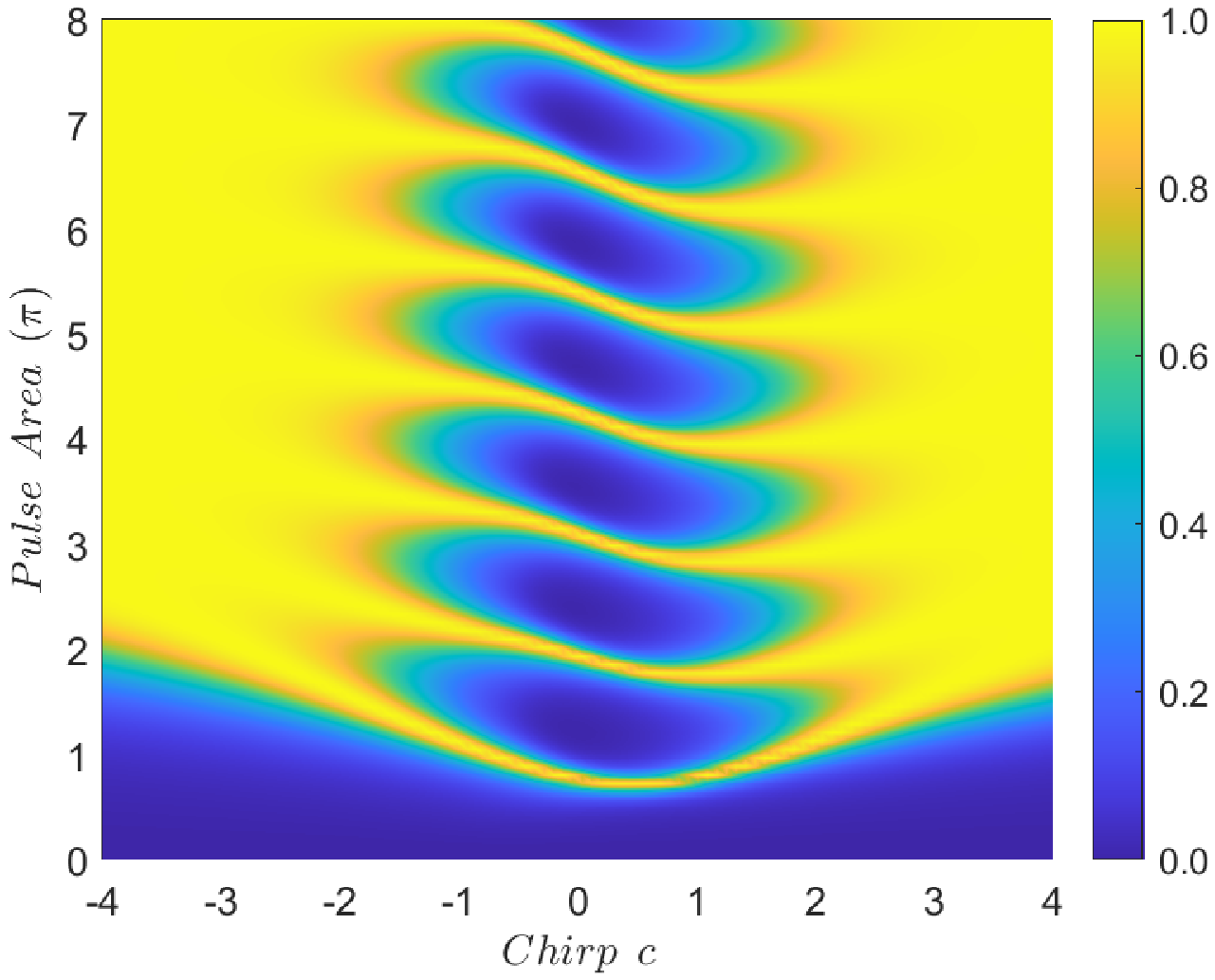}} &
        \subfigure[$\ $]{
	            \label{fig:S_075_12}
	            \includegraphics[width=.45\linewidth]{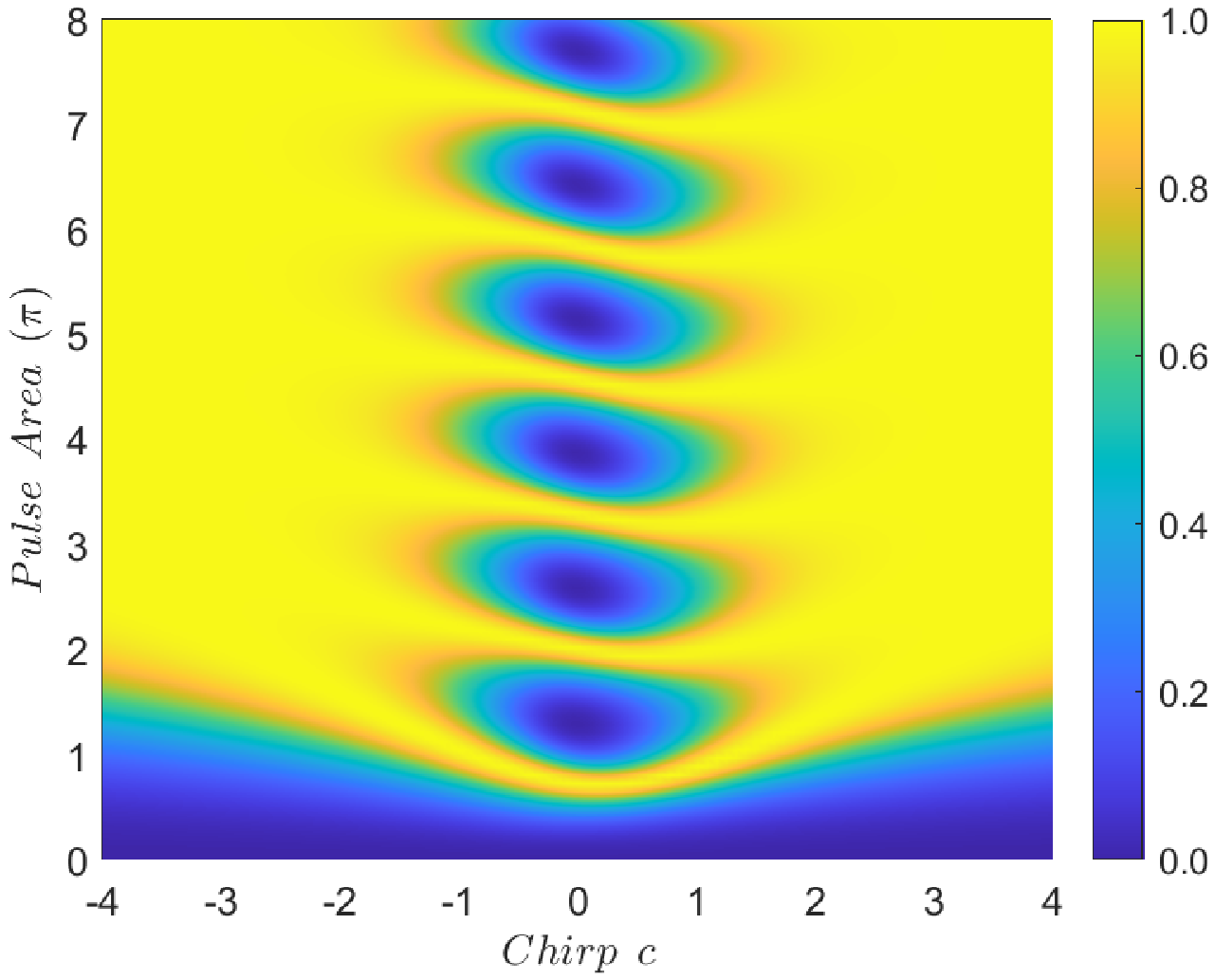}} \\
        \subfigure[$\ $]{
	            \label{fig:S_075_13}
	            \includegraphics[width=.45\linewidth]{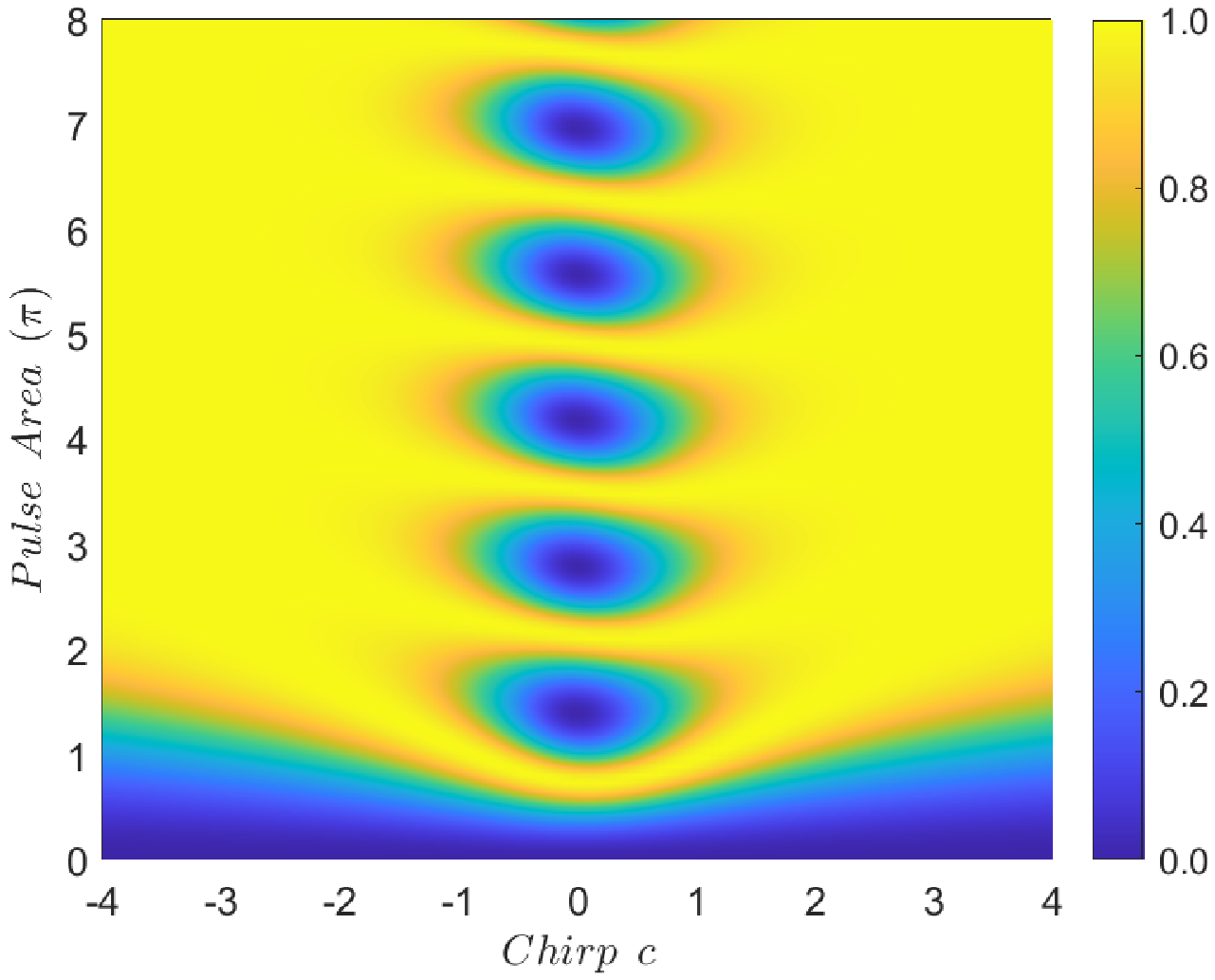}} &
        \subfigure[$\ $]{
	            \label{fig:S_075_15}
	            \includegraphics[width=.45\linewidth]{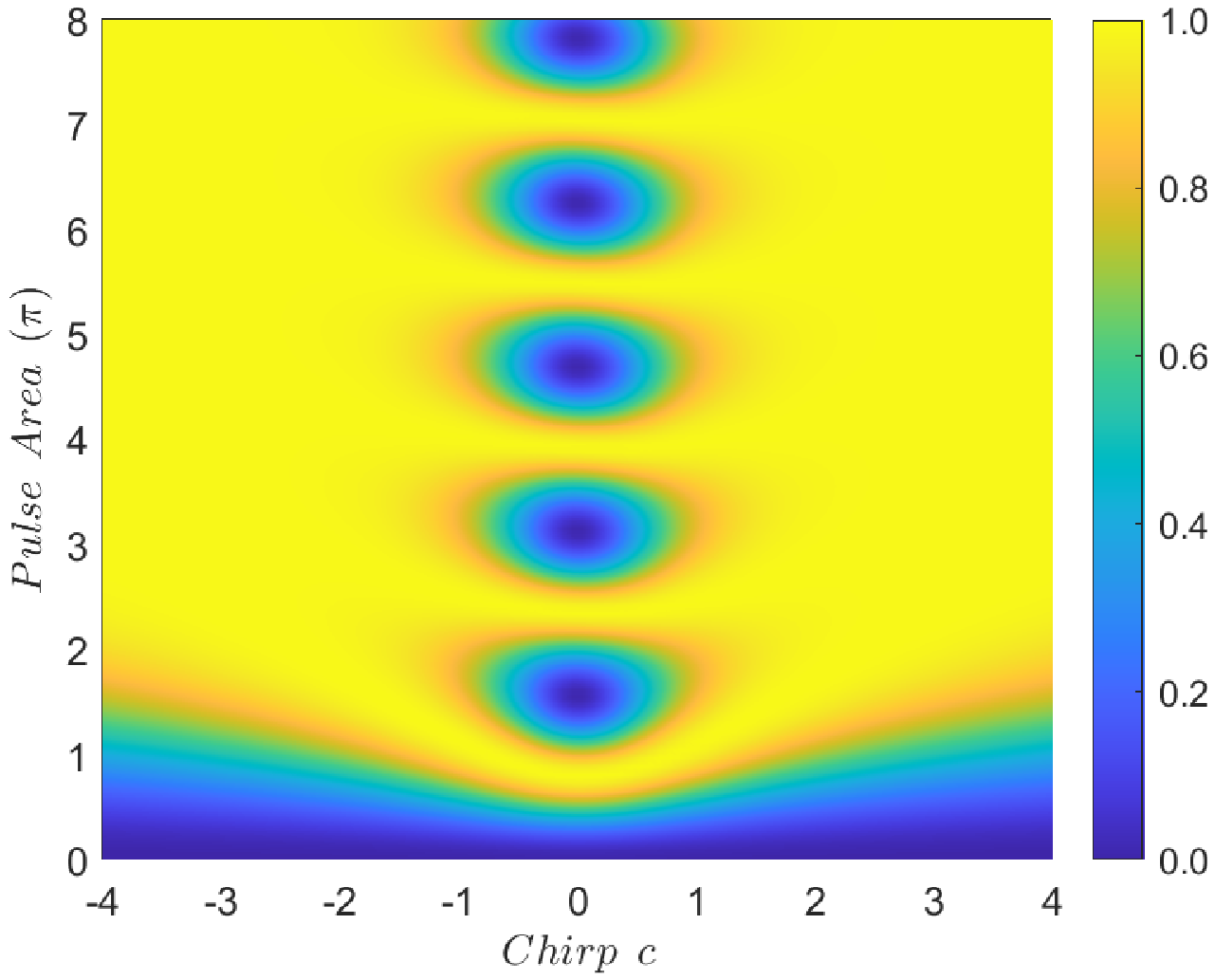}} \\
        \subfigure[$\ $]{
	            \label{fig:S_075_30}
	            \includegraphics[width=.45\linewidth]{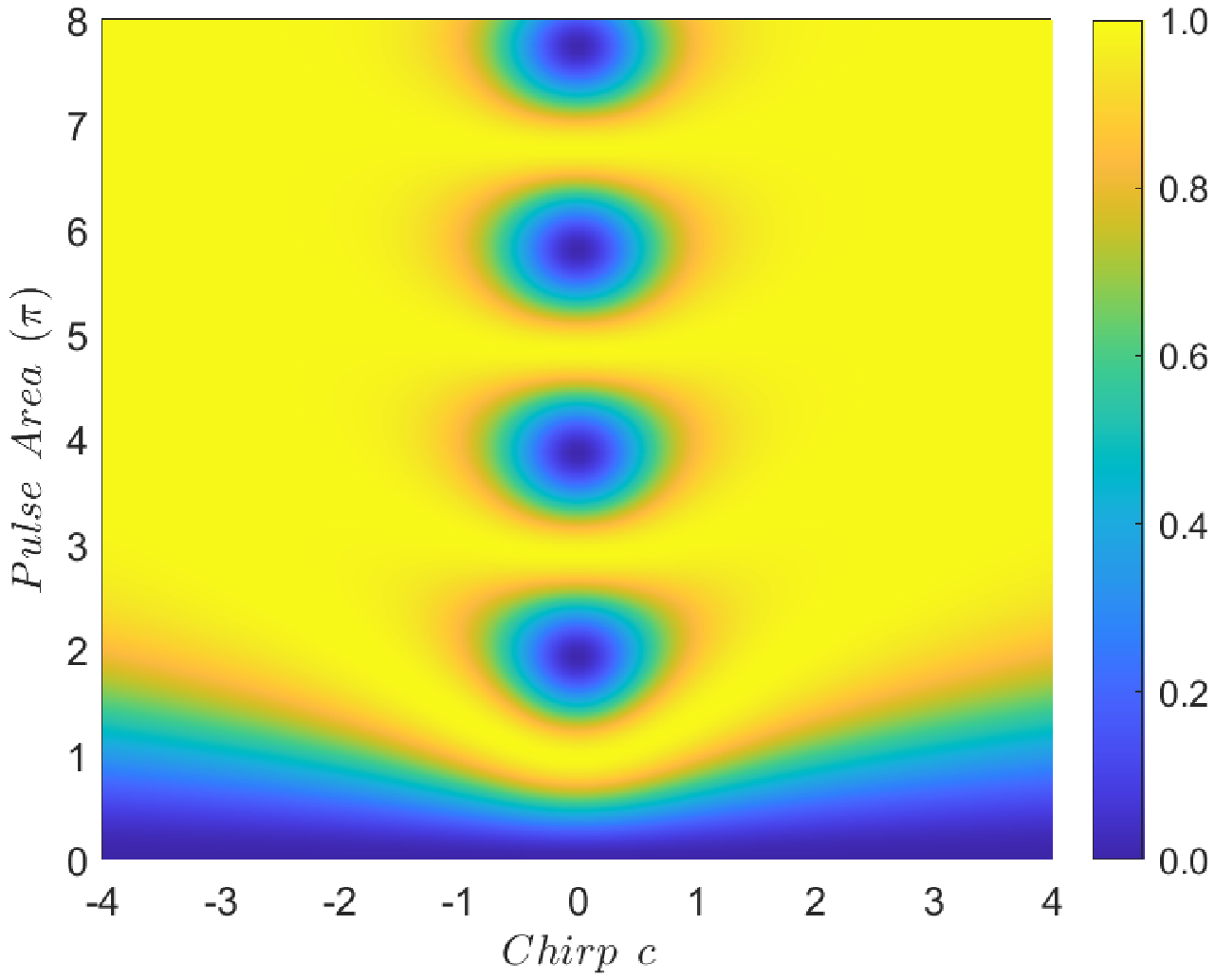}} &
        \subfigure[$\ $]{
	            \label{fig:S_075_80}
	            \includegraphics[width=.45\linewidth]{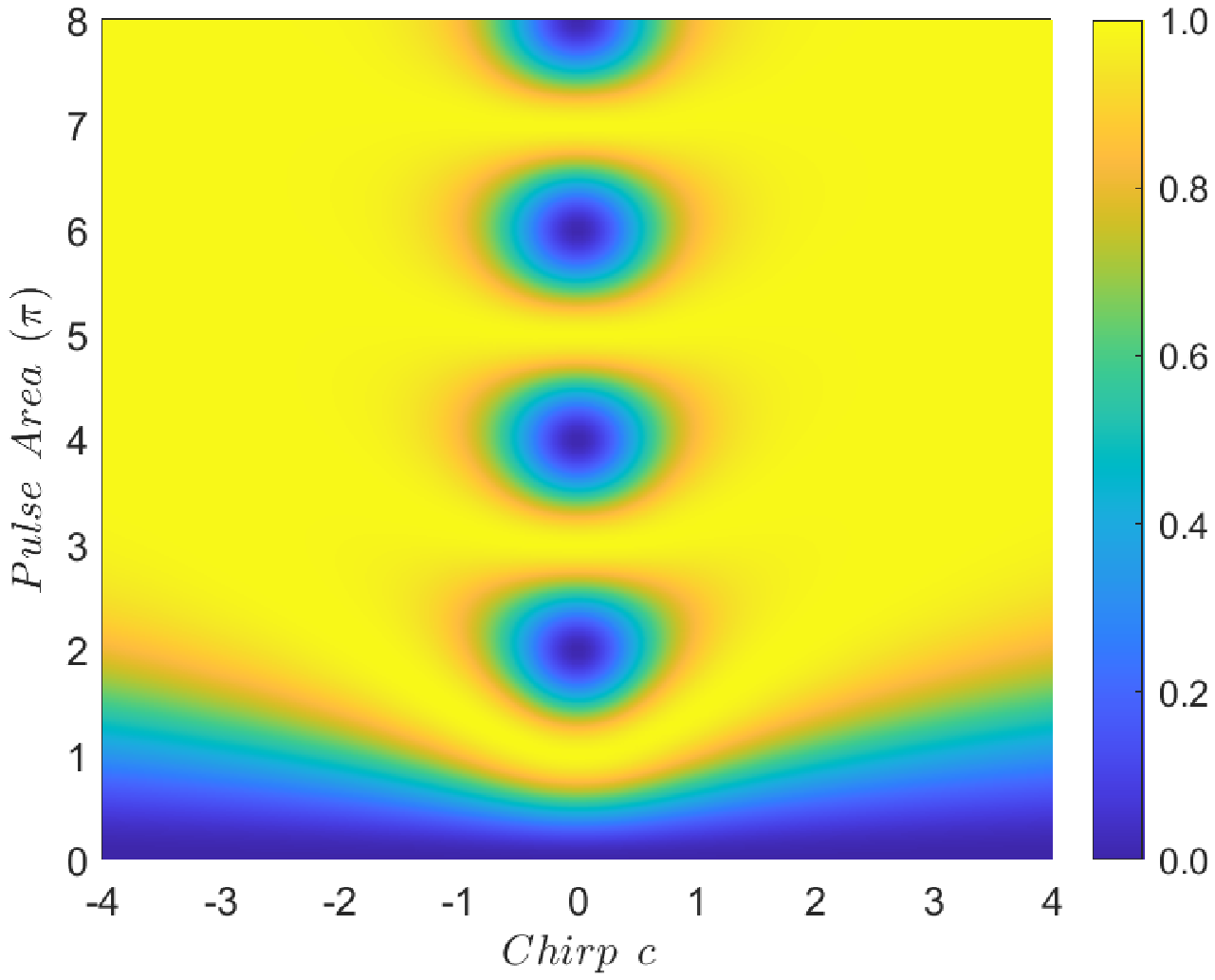}}
		\end{tabular}
\caption{(Color online) Contour plot of the final exciton population when using the hyperbolic secant chirped pulse with $t_p=0.75$ ps, versus the pulse area and the chirp parameter $c$, for different values of the interparticle distance: (a) $R=11$ nm, (b) $R=12$ nm, (c) $R=13$ nm, (d) $R=15$ nm, (e) $R=30$ nm, (f) $R=80$ nm.}
\label{fig:S_075}
\end{figure*}

Here, we test the performance of the previously discussed chirped pulses with numerical simulations of the system Eqs. (\ref{system}), including the effect of nanoparticle as well as relaxation. For the parameters that appear in these equations and are necessary in the simulations, we use numerical values typically corresponding to CdSe-based SQD, which have also been used in many other works regarding similar systems: $T_1 \:=0.8\:$ ns, $T_2\: = 0.3\:$ ns, $\varepsilon_{env}=\varepsilon_0$, $\varepsilon_s=6\varepsilon_0$, $\hbar\omega_0=2.5\:$ eV, $\mu=0.65$ $e$ nm, and $\alpha=7.5\:$ nm, where as usual $\varepsilon_0$ denotes the vacuum dielectric constant. For $\varepsilon_m(\omega)$ we use the experimental value corresponding to gold \cite{Johnson72a}, which is specifically $\varepsilon_m(\omega) = -2.27829 \: + \:i3.81264 $.

In Fig. \ref{fig:G_1} we display contour plots of the final exciton population $\rho_{11}(t_f)$ versus pulse area and chirp, when applying the chirped Gaussian pulse (\ref{G_chirped}) with $\tau_0=1$ ps to the SQD-MNP system, for several interparticle distances. Observe that for the smallest distance $R=11$ nm, Fig. \ref{fig:G_1_11}, efficient population transfer is achieved only at some narrow strips corresponding to specific combinations of pulse area and chirp, resembling the performance of resonant pulses. The reason is that for such small distances the nonlinear term $G$ is very strong and practically destroys the adiabatic following and cancels the beneficial effect of chirp. The situation is drastically improved for a small change in the interparticle distance, see Fig. \ref{fig:G_1_12} corresponding to $R=12$ nm. Due to the reduction of $G$, now the high fidelity stripes become wider and occupy a larger portion of the contour diagram. For distances as small as $13$ nm and $15$ nm, see Figs. \ref{fig:G_1_13} and \ref{fig:G_1_15}, respectively, robust population transfer is accomplished for sufficiently large absolute chirp values and pulse areas larger than a chirp-dependent lower threshold. For larger distances, where the effect of the MNP weakens, the contour diagrams become almost independent of $R$, see f.e. Figs. \ref{fig:G_1_30} and \ref{fig:G_1_80} coresponding to $R=30$ nm and $R=80$ nm, which look identical. Fig. \ref{fig:G_075} is obtained similarly to Fig. \ref{fig:G_1} by applying a linearly chirped Gaussian pulse with shorter $\tau_0=0.75$ ps. Obviously, the overall performance and robustness are now improved. This can be understood by using the analysis of the previous section. Specifically, from Eq. (\ref{eq:ca}) or Fig. \ref{fig:c} we see that, for fixed $a$, the chirp rate $c$ is larger for smaller $\tau_0$. Accordingly, the initial and final chirp values, $|\dot{\phi}(0)|=|\dot{\phi}(2t_0)|=|c|t_0$ are larger for smaller $\tau_0$. But from the discussion at the end of the previous section we know that the unwanted term affecting the dynamics has the form $G_R\Delta(t)$, thus it is stronger at the beginning and at the end, where $|\Delta|$ is close to unity. The larger chirp for smaller $\tau_0$ during the same time intervals, cancels more effectively the undesirable action of this term.  

Probably the most striking feature of Figs. \ref{fig:G_1} and \ref{fig:G_075} is the asymmetry for positive and negative chirp, evident for small values of the interparticle distance, where the symmetry breaking term $G_R\Delta(t)$ identified in the previous section is stronger. In order to numerically confirm the previous theoretical analysis, we manually set $G_R=0$ while keeping nonzero $G_I$ in Eqs. (\ref{system}). In Fig. \ref{fig:GR} we show contour plots of the final exciton population versus pulse area and chirp, for a distance $R=12$ nm and Gaussian chirped pulse with $\tau_0=1$ ps, Fig. \ref{fig:GR_1}, and $\tau_0=0.75$ ps, Fig. \ref{fig:GR_075}. Comparing these to the corresponding figures with nonzero $G_R$, Figs. \ref{fig:G_1_12} and \ref{fig:G_075_12}, it becomes obvious that the asymmetry has been disappeared. We also observe that, if we manually invert the sign of $G_R$, the chirp asymmetry is also inverted, as shown in Fig. \ref{fig:m_GR}, where the same parameters are used as in Figs. \ref{fig:G_1_12} and \ref{fig:G_075_12}. We emphasize of course that these manual changes in $G_R$ are performed only for demonstration reasons, while the real values of this parameter are displayed in Fig. \ref{fig:Re_G}. Another interesting observation which can be made from Figs. \ref{fig:G_1_11}, \ref{fig:G_1_12} and Figs. \ref{fig:G_075_11}, \ref{fig:G_075_12} is that, for negative chirp, once the (larger) pulse area threshold is surpassed and for higher chirp parameter values, the population transfer is more robust compared to positive chirp. This can be explained since for negative chirp the ``detuning" $\dot{\phi}(t)$ and the term $G_R\Delta(t)$ evolve in the same direction, i.e. from positive to negative values, while for positive chirp they evolve in opposite directions (recall that $G_R>0$, see Fig. \ref{fig:Re_G}, while $\Delta(t)$ changes from its maximum value one to negative values). This is also demonstrated in Fig. \ref{fig:gap}, where we display the modified ``eigenvalues" (\ref{modified_eigenvalues}) for positive (red dashed curves) and the opposite negative (blue solid curves) chirp. Specifically, in Fig. \ref{fig:gap_11_1} we plot the ``eigenvalues" for the upper pair of opposite chirp values marked in Fig. \ref{fig:G_1_11}, while in Fig. \ref{fig:gap_11_2} for the lower pair marked in Fig. \ref{fig:G_1_11}. Analogously, in Figs. \ref{fig:gap_12_1}, \ref{fig:gap_12_2} we plot the ``eigenvalues" for the opposite chirp pairs marked in Fig. \ref{fig:G_1_12}. Observe that in all the displayed cases, the gap between the ``eigenvalues" is larger for negative chirp (blue solid curves) than for positive chirp (red dashed curves). We also observe that the difference in the gap for opposite chirp is larger for $R=11$ nm, upper row in Fig. \ref{fig:gap}, than for $R=12$ nm, lower row in Fig. \ref{fig:gap}, since the symmetry breaking parameter $G_R$ is stronger for smaller distances. By manually inverting the sign of $G_R$ this asymmetry is also inverted, as it was demonstrated in Fig. \ref{fig:m_GR}. At this point it is worth to notice that asymmetry in the final exciton population with respect to the chirp sign has also been observed for a SQD without the presence of a MNP, where the main source of decoherence is taken to be the coupling to acoustic phonons \cite{Luker12,Mathew14a}.

In Figs. \ref{fig:S_15} and \ref{fig:S_075} we display contour plots of the final exciton population versus the pulse area and the chirp parameter $c$, when using the hyperbolic secant chirped pulse (\ref{sech}), (\ref{tanh_chirp}) with $t_p=1.5$ ps and $t_p=0.75$ ps, respectively, for the same interparticle distances as in Figs. \ref{fig:G_1} and \ref{fig:G_075}. Despite the differences due to the different type of pulse and chirp, we also observe several similarities with the behavior shown in Figs. \ref{fig:G_1} and \ref{fig:G_075} for linearly chirped Gaussian pulses. Specifically, robust and high-fidelity preparation of the exciton state even for short interparticle distances, when exceeding a chirp dependent pulse area threshold and for larger chirp parameter values, improvement for smaller $t_p$ values, asymmetry in the positive and negative chirp performance for small interparticle distances, and more robust behavior of negative chirp for short distances as long as the pulse area threshold is exceeded. All these observations are explained using similar arguments to those employed for Gaussian linearly chirped pulses.

\section{Conclusion}

\label{sec:con}

In this article, we demonstrated with numerical simulations the efficient generation of the exciton state in a coupled semiconductor quantum dot-metal nanoparticle system, even for short interparticle distances, using conventional chirped pulses with Gaussian and hyperbolic secant profiles. The asymmetry observed in the final exciton population with respect to the chirp sign of the applied pulses was also explained using the system equations. The symmetry breaking term was identified as the real part of the nonlinearity parameter emerging from the interaction between excitons in the quantum dot and plasmons in the metal nanoparticle. The suggested robust quantum control scheme, involving the easily implementable conventional chirped pulses, can find application in the implementation of ultrafast nanoswitches and quantum information processing tasks with semiconductor quantum dots.

\end{document}